\documentclass[12pt]{article}

\usepackage[T1]{fontenc}
\usepackage{lmodern}

\usepackage{setspace}
\usepackage[margin=1.25in]{geometry}

\usepackage[dvipsnames]{xcolor}
\usepackage{pdfpages}
\usepackage{float}
\usepackage{multirow}
\usepackage{array}
\usepackage{caption}
\usepackage{subcaption}
\usepackage{epigraph}

\AtBeginDocument{%
  \everymath{\color{Violet}}%
  \everydisplay{\color{Violet}}%
}

\usepackage{mathtools}
\usepackage{amssymb}
\usepackage{amsthm}
\usepackage{amsfonts}
\usepackage{aliascnt}
\usepackage{accents}
\usepackage{dutchcal}
\usepackage{bbm}

\usepackage{enumitem}
\usepackage{sgame}
\usepackage{tikz}
\usepackage{tikz-cd}
\usetikzlibrary{calc,shapes,arrows}
\usepackage{tcolorbox}
\usepackage{listings}
\usepackage[normalem]{ulem}

\usepackage[round]{natbib}
\DeclareMathOperator*{\argmax}{arg\,max}

\newcommand{\E}{\mathbb{E}}
\newcommand{\Pp}{\mathbb{P}}
\newcommand{\F}{\mathbb{F}}
\newcommand{\G}{\mathbb{G}}
\newcommand{\calF}{\mathcal{F}}
\newcommand{\calG}{\mathcal{G}}
\newcommand{\calH}{\mathcal{H}}
\newcommand{\calE}{\mathcal{E}}
\newcommand{\dd}{\,\mathrm{d}}
\newcommand{\Law}{\mathcal{L}}
\newcommand{\icx}{\preceq_{\mathrm{icx}}}
\newcommand{\R}{\mathbb{R}}
\newcommand{\Q}{\mathbb{Q}}

\newcommand{\1}{\mathbbm{1}}

\newtheorem{theorem}{Theorem}[section]

\newaliascnt{proposition}{theorem}
\newtheorem{proposition}[proposition]{Proposition}
\aliascntresetthe{proposition}

\newaliascnt{lemma}{theorem}
\newtheorem{lemma}[lemma]{Lemma}
\aliascntresetthe{lemma}

\newaliascnt{corollary}{theorem}
\newtheorem{corollary}[corollary]{Corollary}
\aliascntresetthe{corollary}

\newaliascnt{claim}{theorem}
\newtheorem{claim}[claim]{Claim}
\aliascntresetthe{claim}

\theoremstyle{definition}

\newaliascnt{definition}{theorem}
\newtheorem{definition}[definition]{Definition}
\aliascntresetthe{definition}

\newaliascnt{property}{theorem}

\aliascntresetthe{property}

\newaliascnt{example}{theorem}
\newtheorem{example}[example]{Example}
\aliascntresetthe{example}

\newaliascnt{assumption}{theorem}
\newtheorem{assumption}[assumption]{Assumption}
\aliascntresetthe{assumption}

\newaliascnt{condition}{theorem}

\aliascntresetthe{condition}

\newaliascnt{question}{theorem}

\aliascntresetthe{question}

\newaliascnt{remark}{theorem}
\newtheorem{remark}[remark]{Remark}
\aliascntresetthe{remark}

\newaliascnt{remarks}{theorem}

\aliascntresetthe{remarks}

\newaliascnt{aside}{theorem}

\aliascntresetthe{aside}

\newaliascnt{note}{theorem}

\aliascntresetthe{note}

\usepackage{thmtools}
\usepackage{thm-restate}

\usepackage{hyperref}

\hypersetup{
    colorlinks=true,
    linkcolor=OrangeRed,
    filecolor=Thistle,
    urlcolor=Thistle,
    citecolor=Thistle,
}

\usepackage[nameinlink]{cleveref}

\crefname{theorem}{theorem}{theorems}
\Crefname{theorem}{Theorem}{Theorems}
\crefname{proposition}{proposition}{propositions}
\Crefname{proposition}{Proposition}{Propositions}
\crefname{lemma}{lemma}{lemmas}
\Crefname{lemma}{Lemma}{Lemmas}
\crefname{corollary}{corollary}{corollaries}
\Crefname{corollary}{Corollary}{Corollaries}
\crefname{claim}{claim}{claims}
\Crefname{claim}{Claim}{Claims}
\crefname{definition}{definition}{definitions}
\Crefname{definition}{Definition}{Definitions}
\crefname{property}{property}{properties}
\Crefname{property}{Property}{Properties}
\crefname{example}{example}{examples}
\Crefname{example}{Example}{Examples}
\crefname{assumption}{assumption}{assumptions}
\Crefname{assumption}{Assumption}{Assumptions}

\makeatletter
\let\cref@old@isrefconsecutive\cref@isrefconsecutive
\def\cref@isrefconsecutive#1#2{%
  \begingroup
    \def\cref@assumptiontype{assumption}%
    \cref@gettype{#1}{\cref@typea}%
    \ifx\cref@typea\cref@assumptiontype
      \endgroup
      \@cref@refconsecutivefalse
    \else
      \endgroup
      \cref@old@isrefconsecutive{#1}{#2}%
    \fi
}
\makeatother

\crefname{condition}{condition}{conditions}
\Crefname{condition}{Condition}{Conditions}
\crefname{question}{question}{questions}
\Crefname{question}{Question}{Questions}
\crefname{remark}{remark}{remarks}
\Crefname{remark}{Remark}{Remarks}
\crefname{remarks}{remarks}{remarks}
\Crefname{remarks}{Remarks}{Remarks}
\crefname{aside}{aside}{asides}
\Crefname{aside}{Aside}{Asides}
\crefname{note}{note}{notes}
\Crefname{note}{Note}{Notes}
\crefname{appendix}{appendix}{appendices}
\Crefname{appendix}{Appendix}{Appendices}

\newcommand{\secref}[1]{\hyperref[#1]{\S\ref*{#1}}}

\definecolor{backcolour}{rgb}{0.63,0.79,0.95}
\lstdefinestyle{mystyle}{
  backgroundcolor=\color{backcolour},
  basicstyle=\ttfamily\footnotesize,
  breakatwhitespace=false,
  breaklines=true,
  captionpos=b,
  keepspaces=true,
  numbers=left,
  numbersep=5pt,
  showspaces=false,
  showstringspaces=false,
  showtabs=false,
  tabsize=2
}
\begin{document}
\title{Best Garbling is No Garbling: Persuasion in Real Time}
\author{Can Urgun\thanks{UNC Chapel Hill. Email: \href{mailto:curgun@unc.edu}{curgun@unc.edu}.} \and Mark Whitmeyer\thanks{Arizona State University. Email: \href{mailto:mark.whitmeyer@gmail.com}{mark.whitmeyer@gmail.com}. We thank seminar and conference audiences at UNC/Duke, Iowa, and SITE for their invaluable feedback. We used ChatGPT in the writing and development of this paper and solicited feedback from \href{refine.ink}{refine.ink}.}}
\date{September 2026}

\maketitle

\begin{abstract}
We study continuous-time persuasion where a sender controls both how informative a signal is over time and when to stop providing information to a receiver. There is an exogenous signal process, which the sender both garbles and stops, delaying the receiver's decision at a convex, increasing cost of time. We show that, although both instruments are available, under natural regularity conditions, \textit{full transparency is always optimal}: the sender keeps the signal fully informative and persuades solely by choosing when to stop.
\end{abstract}

\section{Introduction}\label{sec:introduction}

Many environments--drug approvals, loan underwriting, jury verdicts, hiring and tenure votes, content moderation, even ``buy/not-buy'' choices--share the same essential features: one agent assembles and releases evidence to sway another agent into taking a favorable binary decision, which takes a cut-off form. For instance, a regulator or internal evaluator approves a drug being tested if and only if her belief that the drug works exceeds a cutoff; a capital-provider funds a loan if and only if default risk falls below a hurdle; and a median juror votes to convict if and only if the guilt probability passes a ``reasonable doubt'' or ``preponderance of evidence'' threshold. 

The canonical persuasion problem \citep{kamenica2011bayesian} captures this logic in one sentence: a sender, facing a binary state and a receiver with a threshold rule for a binary action, commits to an information structure to maximize the chance the receiver's posterior exceeds the threshold. What the classic model leaves out, and what real cases make unavoidable, is that moving beliefs takes time: evidence must be generated, filtered, and digested, and delay is costly.

In our setting, we keep this threshold-persuasion objective as our primary focus but make the information flow and the stopping time explicit--signals arrive over time, the receiver updates continuously and acts upon stopping, and the sender pays a convex cost of delay--so the design problem becomes: how should a sender shape and pace what is revealed to persuade a receiver \textit{at the lowest (expected) cost}?

More concretely, the sender and the receiver share a common prior about an unknown binary state that drives a Brownian process \(X\) in continuous time. The receiver does not observe the exogenous source directly, however, but through a sender-chosen information structure that garbles \(X\). The sender chooses both the informativeness of the real-time disclosure and a stopping time at which the receiver acts, incurring a convex increasing cost of delay. Upon stopping, the receiver takes a binary action, choosing the sender-preferred action if and only if her belief exceeds a threshold.

To answer our primary question (solving the threshold-objective problem), we solve a much more general one. Our main result says that for a broad class of policies,\footnote{The sender has neither advance information about future evidence nor any state information beyond the evidence already observed (\Cref{ass:binary-brownian-evidence}). In addition, the policy must satisfy a property we call clock-dominance (\Cref{def:clock-domination}): at every date \(t\), the receiver's posterior equals \(Q_s\) for some \(s\leq t\), where \(Q\) is a copy of the sender's posterior process (and, hence, of the posterior the receiver would hold under full transparency).} no matter the (implementable) target distribution over posteriors the sender wishes to implement via garbling and stopping and no matter the time cost (so long as it is convex), the ``best garbling is no garbling.'' That is, \Cref{thm:clock-bgng} reveals that there is a solution to the sender's persuasion problem in which she does not garble the exogenous process at all and influences the receiver's behavior through the choice of stopping time alone. The basic intuition for this is that garbling has a net \textit{time-risk-increasing} effect on the receiver's belief process, which is suboptimal given the convex time cost.

As a first step, we show that for any implementable terminal posterior distribution, no public policy can produce it in less expected time than full transparency, and every transparent stopping rule that produces it has the same expected duration (\Cref{prop:mean-time-efficiency} and \Cref{cor:transparent-mean-time}). Expected duration alone does not settle the problem under a convex cost, however--the shape of the distribution over times at which the sender stops is crucial.

To tackle this problem, we introduce the concept of a \textit{disclosure clock}, which allows us to rewrite the public posterior process \(\widehat P\) produced by the sender as a time change of a copy of the sender's posterior process (which the receiver would also hold under full transparency). That is, the disclosure clock \(C_t\) measures how long this copy takes to reach the public posterior at date \(t\): \(Q_{C_t}=\widehat P_t\) for all \(t\). Our central assumption, then, is that the sender's policy is clock-dominated; \textit{viz.}, \(C_t \leq t\) almost surely, for all \(t\). There is no analog of this restriction in a static model: garbling already makes the receiver's posterior a mean-preserving contraction of the sender's posterior, whereas clock domination requires these contractions to be compatible across dates, so that public learning never runs ahead of full transparency. This rules out rapid, we contend pathological, release of information by the sender and facilitates the chain of inequalities we ultimately use to prove the result.

Having shown that full transparency is without loss of optimality, we study the transparent stopping problem for the threshold objective. The comparative statics are intuitive: higher incremental delay costs lower the corresponding minimum and maximum persuasion probabilities (\Cref{prop:higher-delay-cost-lower-persuasion}). In turn, a higher signal-to-noise ratio raises the persuasion probabilities (\Cref{prop:higher-snr-higher-persuasion}).

We finish by studying an extension of the threshold-objective problem where the receiver can also choose when to stop. There, our argument for \Cref{thm:clock-bgng} does not apply, because the receiver may prefer to stop earlier. However, under a slightly stronger (Lipschitz) condition, every equilibrium has a transparent acceleration that remains optimal for the receiver and benefits both players (\Cref{thm:uncommitted-receiver-acceleration}).

\secref{sec:model} presents the environment. \secref{sec:bgng} develops the disclosure-clock comparison, and \secref{sec:mainresult} proves the main theorem. \secref{sec:cs} establishes the comparative statics, and \secref{sec:uncommitted-receiver} studies the receiver-stopping extension. All proofs are collected in the appendices.

\subsection{Related Work}\label{sec:relatedwork}

A growing literature studies dynamic persuasion and information design. Relative to this work and the static benchmark \citep{kamenica2011bayesian}, our contribution is to make the tempo of information part of the designer's choice. Methodologically, we bring Brownian methods into persuasion by allowing the sender to choose both how transparent to be about a continuous-time diffusion and when to stop.

Several papers study dynamic persuasion with costly or gradual information, but treat time differently.\footnote{\citet{CheKimMierendorff2023Dynamic} analyze dynamic persuasion when information takes time to generate and neither player can commit. As persuasion costs vanish, Markov equilibria approximate both the \citet{kamenica2011bayesian} optimum and full revelation, but the sender has no separate stopping instrument. \citet{EscudeSinander2023Slow} study ``slow persuasion'' under a gradual-accumulation constraint and characterize when the sender benefits from slowing down learning. Other recent work combines dynamic disclosure with additional strategic forces like multiple audiences or incentive provision. See, e.g., \citet{LiSzydlowskiYu2025Entry} and \citet{ElyGeorgiadisRayo2025Feedback}.} Our continuous-time (Brownian) setting is related to \citet{AybasCallander2024CheapTalk}, who use Brownian motion to represent complexity and study when one-shot cheap talk can be efficient, and to work comparing cheap talk and persuasion, such as \citet{KamenicaLin2024Commitment}.\footnote{On the static side, papers that exploit the geometric structure of persuasion problems include \citet{BardhiGuo2018Modes} and \citet{GuoShmaya2019Interval}.} Closest in spirit are models where influence comes from controlling the continuation of an evidence stream.\footnote{\citet{brocas_carrillo_2007_influence} study a discrete-time setting in which a sender gains influence by deciding when to stop generating public evidence (``influence through ignorance''). \citet{henry2019research} explore a continuous-time approval environment in which an informer pays a flow cost to run a fixed Brownian experiment and an evaluator decides whether to approve or reject. Starting from Wald's sequential test \citep{wald1945sequential}, they analyze how organizational form distorts stopping (see also \citet{McClellan2022}). Our paper tackles a different design problem: rather than fixing the experiment and varying who controls stopping, we take the continuous-time information technology itself as an object of design, allowing the sender to choose transparency as well as when to stop.} Like ours, both \citet{OrlovSkrzypaczZryumov2020} and \citet{ArieliBabichenkoShaidermanShi2025} provide conditions in which full transparency (perhaps with delay) is optimal.\footnote{The former study a real-options problem in which a biased sender controls disclosure while the receiver chooses when to act and show that delayed full disclosure is optimal because it persuades the receiver to wait. The latter study a discrete-time product-adoption problem in which the sender learns privately over time and show that, when the sender is sufficiently impatient and her belief follows a random walk, full transparency is optimal until adoption.}

Another strand combines persuasion with search and outside information sources.\footnote{\citet{BizzottoRudigerVigier2021Outside} study dynamic persuasion when the receiver also observes outside news and can stop listening, while \citet{MekonnenMurraAntonPakzadHurson2025PersuadedSearch} and \citet{MekonnenPakzadHurson2025Competition} analyze information brokers who design signals for a searching agent.} On the information-acquisition side, \citet{Zhong2022ODIA} and \citet{ChenZhong2025TimeRisk} study single-agent dynamic problems in which a decision maker chooses a signal process subject to constraints or costs and trades off delay against accuracy, while \citet{BloedelZhong2020Cost} characterize which posterior-separable reduced-form costs arise from optimal sequential sampling. \citet{MorrisStrack2019} show that, in the two-state Wald problem, any distribution of posteriors can be implemented by a stopping rule and the induced ex ante cost is posterior-separable.\footnote{For related sequential-sampling models of response times and learning in rich environments, see, e.g., \citet{FudenbergStrackStrzalecki2018,Goncalves2022SSE,Goncalves2024SAC,Barilla2025WhenWhat,GeorgiadisHarrisPreparingToAct}.}

\section{Model}\label{sec:model}

There is an unknown state, \(\mu\in\{\mu_h,\mu_\ell\}\), where \(\mu_h>\mu_\ell\). A sender and a receiver share a full-support prior \(p_0\coloneqq\Pp(\mu=\mu_h) \in (0,1)\). The unknown state is the drift of an evidence process \(X\), which evolves according to
\[
\dd X_t=\mu \dd t+\sigma \dd B_t,
\]
where \(\sigma>0\) is known and \(B\) is a Brownian motion independent of \(\mu\).\footnote{We use \(B\) for the noise in the evidence process and \(W\) for Brownian motions driving posterior processes.} The ratio \(\kappa\coloneqq(\mu_h-\mu_\ell)/\sigma\) measures the signal-to-noise ratio. Let \(\F^X=(\calF_t^X)_{t\geq0}\) be the usual augmentation of the natural filtration of \(X\).\footnote{The natural filtration records the history of the process. A filtration is \textit{usual} if it is complete and right-continuous; its usual augmentation adds null events and imposes right-continuity.}

The sender may randomize her reports, and a report may contain noise that is independent of the state. Accordingly, we allow the ambient filtration \(\F\) to contain auxiliary randomness in addition to the evidence history \(\F^X\). Our next assumption ensures that this additional randomness neither reveals future evidence nor supplies another signal about the state.
\begin{assumption}\label{ass:binary-brownian-evidence}
Let \(\F=(\calF_t)_{t\geq0}\) be a usual filtration containing \(\F^X\), and let \(\Pp^\ell\) and \(\Pp^h\) be the conditional laws under the low and high states. Under \(\Pp^\ell\), assume \(B_t^\ell\coloneqq(X_t-\mu_\ell t)/\sigma\) is an \(\F\)-Brownian motion. For every finite \(t\), assume
\[\left.\frac{\dd\Pp^h}{\dd\Pp^\ell}\right|_{\calF_t} =Z_t \coloneqq\calE(\kappa B^\ell)_t =\exp\left(\kappa B_t^\ell-\frac{\kappa^2}{2}t\right).\]
\end{assumption}
\Cref{ass:binary-brownian-evidence} is mild, imposing two natural restrictions on the information in \(\F\). First, requiring \(B^\ell\) to remain an \(\F\)-Brownian motion prevents \(\calF_t\) from revealing future increments of the evidence process. Intuitively, the sender does not have ``access to the future.'' Second, the likelihood-ratio formula says that the evidence history contains all of the state information in \(\calF_t\). In particular, for every \(A\in\calF_t\), \(\Pp^h\left(A\mid\calF_t^X\right) = \Pp^\ell\left(A\mid\calF_t^X\right)\) almost surely. In short, the sender may use the additional randomness in \(\F\) to randomize or add noise to her reports, but that randomness cannot reveal anything about the state once the evidence history is known. 

Girsanov's theorem also produces the evidence dynamics under the high state. Under \(\Pp^h\),
\(B_t^h\coloneqq B_t^\ell-\kappa t\) is an \(\F\)-Brownian motion, and therefore
\(X_t=\mu_h t+\sigma B_t^h\). Consequently, observing all of \(\calF_t\) produces the same posterior as observing the evidence history alone:
\[
\Pp\left(\mu=\mu_h\mid\calF_t\right)
=
\Pp\left(\mu=\mu_h\mid\calF_t^X\right)
=
\frac{p_0Z_t}{1-p_0+p_0Z_t}.
\]
Because \(\Pp^h\) and \(\Pp^\ell\) are the laws conditional on the high and low states, the law of total probability yields, for every event \(A\),
\(\Pp(A)=p_0\Pp^h(A)+(1-p_0)\Pp^\ell(A)\). We go through a concrete microfoundation for reporting technologies satisfying \Cref{ass:binary-brownian-evidence} in \Cref{app:microfound}. In particular, we allow for full transparency and continuous reports; coarse labels; silence, delayed disclosure, and release of information in batches; and policies adding both continuous and (Poisson) jump noise.

The sender observes \(X\) in real time. The receiver observes only a committed public report and forms a posterior \(\widehat P\) from its public history. At the sender's public stopping time \(\tau\), the receiver chooses an action in \(\{0,1\}\). There is a threshold \(\bar P\in(p_0,1)\) such that the receiver takes the sender-preferred action if and only if \(\widehat P_\tau\geq\bar P\). The sender receives one when that action is taken and zero otherwise. Her delay cost \(c\colon[0,\infty)\to[0,\infty)\) is finite, continuous, nondecreasing, convex, and satisfies \(c(0)=0\). Her payoff is
\[\E\left[\1_{\{\widehat P_\tau\geq\bar P\}}-c(\tau)\right].\]
We restrict attention to almost surely finite policies satisfying \(\E[\tau]<\infty\).

\section{Disclosure Clocks and Transparent Persuasion}\label{sec:bgng}

A public experiment specifies the information revealed to both players over time. Formally, it is a usual subfiltration \(\G=(\calG_t)_{t\geq0}\) of \(\F\), where \(\calG_t\) records the public report history through date \(t\). A public policy consists of a public experiment and an almost surely finite \(\G\)-stopping time \(\tau\). Write \(\widehat P_t\) for the receiver's posterior at date \(t\).

\paragraph{Full transparency} Under full transparency, the receiver observes the entire evidence history. Let \(P_t\coloneqq\Pp(\mu=\mu_h\mid\calF_t)\) be the resulting posterior. Because the sender observes \(X\) in real time, \(P\) is also the \textit{sender's own} posterior. We note the following easy facts:

\begin{remark}
Maintain \Cref{ass:binary-brownian-evidence}. There is an \(\F\)-Brownian motion \(W\) under \(\Pp\) such that \[\label{eq:speedoftrans}\tag{1}\dd P_t=\kappa P_t(1-P_t) \dd W_t, \qquad \text{with } P_0 = p_0.\] Moreover,
\[\label{eq:samefilt}\tag{2}\log\frac{P_t}{1-P_t}=\log\frac{p_0}{1-p_0} + \frac{\mu_h-\mu_\ell}{\sigma^2}
\left(X_t-\frac{\mu_h+\mu_\ell}{2}t\right).
\]
Consequently, \(P_t\in(0,1)\) at every finite time, and the natural filtrations of \(P\) and \(X\) coincide up to completion.
\end{remark}

\eqref{eq:speedoftrans} identifies the natural speed of transparent learning. When the posterior is \(p\), its instantaneous volatility is \(\kappa p(1-p)\); that is, a higher signal-to-noise ratio makes beliefs move faster. Moreover, the histories of \(P\) and \(X\) contain the same information. A transparent stopping rule can consequently be written using either process.

\subsection{Expected Time to Produce a Terminal Posterior Law}\label{subsec:mean-time-efficiency}

We now ask how much calendar time a public policy must take on average to produce a specified terminal distribution over posteriors. In particular, we derive a lower bound on \(\E[\tau]\) for every public policy that produces the distribution. We then show that, under full transparency, every integrable stopping rule that produces the specified terminal posterior distribution attains this lower bound.

For a public experiment \(\G=(\calG_t)_{t\geq0}\), define its public likelihood ratio and corresponding posterior by
\[
\Lambda_t\coloneqq\left.\frac{\dd\Pp^h}{\dd\Pp^\ell}\right|_{\calG_t}
\qquad \text{and} \qquad
\widehat P_t\coloneqq\frac{p_0\Lambda_t}{1-p_0+p_0\Lambda_t}.
\]
For probability measures \(\mathsf P\ll\mathsf Q\), write
\[
\mathrm D(\mathsf P\Vert\mathsf Q)
\coloneqq\int\log\left(\frac{\dd\mathsf P}{\dd\mathsf Q}\right) \dd\mathsf P,
\]
for relative entropy, which measures how well the public history distinguishes one state from the other. Define
\[
\Psi(p)\coloneqq(2p-1)\log\frac{p}{1-p},
\qquad
\text{for } p\in(0,1).
\]
For a public policy that stops at \(\tau\), the quantity \(\E[\Psi(\widehat P_\tau)]-\Psi(p_0)\) equals \(p_0\) times the relative entropy from the distribution of the public report history under the high state to its distribution under the low state, plus \(1-p_0\) times the relative entropy in the opposite direction. We say that a stopping time \(S\) \textit{embeds} a probability distribution \(\nu\) in a posterior process if the posterior at \(S\) has law \(\nu\).
\begin{proposition}\label{prop:mean-time-efficiency}
Maintain \Cref{ass:binary-brownian-evidence}. Let \(\tau\) be an almost surely finite \(\G\)-stopping time satisfying \(\E[\tau]<\infty\). Then
\[
\E[\tau]\geq\frac{2}{\kappa^2}\left(\E[\Psi(\widehat P_\tau)]-\Psi(p_0)\right).
\]
Let \(Q\) be any solution in a usual filtration to
\[\label{eq:full-speed-posterior}\tag{3}
\dd Q_t=\kappa Q_t(1-Q_t)\dd W_t,
\qquad \text{with }
Q_0=p_0.
\]
If \(S\) is a stopping time satisfying \(\E[S]<\infty\), then \(\E[S]=\frac{2}{\kappa^2}\left(\E[\Psi(Q_S)]-\Psi(p_0)\right)\).\end{proposition}
We call any process satisfying \eqref{eq:full-speed-posterior} a \textit{full-speed posterior}. The full-transparency process \(P\) satisfies the stochastic differential equation specified in the second part of \Cref{prop:mean-time-efficiency}. Because \(P\) is bounded, its stopped process is uniformly integrable. Thus, applying that part of the proposition to \(P\) reveals that every integrable stopping time embedding \(\nu\) in \(P\) has the same expectation. The first part of the proposition shows that every public policy producing \(\nu\) has expected duration greater than this common expectation. Formally,

\begin{corollary}\label{cor:transparent-mean-time}
Maintain \Cref{ass:binary-brownian-evidence}. Fix a distribution \(\nu\) on \((0,1)\) that can be embedded in \(P\) by an integrable stopping time. Every integrable stopping time that embeds \(\nu\) in \(P\) has the same expectation. Moreover, if a public policy stops at \(\tau\), where \(\E[\tau]<\infty\) and \(\Law(\widehat P_\tau)=\nu\), then its expected duration is greater than this common expectation.
\end{corollary}

The corollary settles the comparison of means: holding the terminal posterior distribution fixed, every transparent stopping rule that produces it has the same expected duration, and no public policy can produce it in less expected time. The first part of the corollary is the constant-flow-cost case of \citet[Propositions~2-3]{MorrisStrack2019}. Their formula continues to hold in our setting because the additional randomization allowed by \Cref{ass:binary-brownian-evidence} does not change the evolution of the posterior under full transparency.

\subsection{Timing Risk}\label{subsec:root-optimality}

Expected time is not enough when delay costs are convex. Two stopping times may have the same mean but different dispersion, and the more dispersed time is more costly under every strictly convex delay cost. We use the convex and the increasing convex orders to make this precise.

\begin{definition}\label{def:increasing-convex-order}
For nonnegative integrable random variables \(S\) and \(T\), write \(S\icx T\) if \(\E[\phi(S)]\leq\E[\phi(T)]\) for every increasing convex function \(\phi\) for which the expectations exist. Write \(S\preceq_{\mathrm{cx}}T\) if the same inequality holds for every convex \(\phi\).\footnote{That is to say, \(S\preceq_{\mathrm{cx}}T\) if and only if
\(S\icx T\) and \(\E[S]=\E[T]\).}
\end{definition}

\citet{root1969} introduces a stopping rule based on a deterministic barrier in the plane with time on one axis and the posterior on the other.\footnote{A Root barrier is a closed set \(\mathcal B\subseteq[0,\infty)\times(0,1)\) such that, if \((t,p)\in\mathcal B\), then \((s,p)\in\mathcal B\) for every \(s\geq t\). Thus, at each posterior \(p\), the barrier specifies a cutoff date, possibly infinite. The rule stops when the current posterior is \(p\) at or after that date. The barrier is deterministic because these cutoff dates are fixed before the process begins.} For a full-speed posterior process \(Q\), write \(R_{\mathcal B}(Q)\coloneqq\inf\{t\geq0\colon(t,Q_t)\in\mathcal B\}\) for the first date at which the current date and posterior lie in \(\mathcal B\).

Our next proposition combines \citet[Theorems~2.8 and~3.3]{CoxOblojTouzi2019} with the average identity in \Cref{prop:mean-time-efficiency}.

\begin{proposition}[Root's timing comparison]\label{prop:root-posterior-timing}
Maintain \Cref{ass:binary-brownian-evidence}. Let \(Q\) be a \([0,1]\)-valued process adapted to a usual filtration, and suppose that \(Q\) satisfies the stochastic differential equation \eqref{eq:full-speed-posterior} with respect to a Brownian motion in that filtration. Let \(\nu\) be a probability distribution on \((0,1)\), and suppose that some integrable stopping time embeds \(\nu\) in \(Q\). Then there is a deterministic Root barrier \(\mathcal B^\nu\) such that \(R_{\mathcal B^\nu}(Q)\) is integrable, embeds \(\nu\) in \(Q\), and satisfies \(R_{\mathcal B^\nu}(Q)\preceq_{\mathrm{cx}}S\) for every integrable stopping time \(S\) that embeds \(\nu\) in \(Q\).
\end{proposition}

Because \(P\) and \(Q\) are both full-speed posteriors, they have the same law. Fix a distribution \(\nu\), and suppose that an integrable stopping time \(S\) embeds \(\nu\) in \(P\). Root's barrier produces a first-entry time \(R^\nu\coloneqq R_{\mathcal B^\nu}(P)\) that also embeds \(\nu\) in \(P\). In the form stated in \Cref{lem:root-icx}, \citet{CoxOblojTouzi2019} show that \(R^\nu\icx S\). By \Cref{prop:mean-time-efficiency}, \(R^\nu\) and \(S\) have the same mean because they embed the same terminal posterior distribution in \(P\). Together, these two results imply \(R^\nu\preceq_{\mathrm{cx}}S\) and so, among the stopping times that embed \(\nu\) in \(P\), Root's construction preserves expected duration and reduces dispersion.

To apply \Cref{prop:root-posterior-timing} to a public policy, it remains to express the policy's posterior on the operating-time scale of a full-speed posterior.

\section{Main Result}\label{sec:mainresult}

\subsection{Clock Domination}\label{subsec:information-clock}

A public report determines a posterior process \(\widehat P\). We compare its timing with full transparency by writing \(\widehat P_t=Q_{C_t}\), where \(Q\) is a full-speed posterior. The clock \(C_t\) measures how long \(Q\) takes to reach the public posterior at date \(t\).
\begin{definition}\label{def:clock-domination}
Let \(\G\) be a public experiment. A \textit{disclosure clock} for \(\G\) is an increasing right-continuous \(\G\)-adapted process \(C=(C_t)_{t\geq0}\), with \(C_0=0\), for which, after enlarging the probability space by independent auxiliary randomness if necessary, there exist a usual filtration \(\mathbb J=(\mathcal J_u)_{u\geq0}\) and a \(\mathbb J\)-adapted full-speed posterior \(Q\) such that:
\begin{enumerate}[label=(\roman*),noitemsep,nosep,leftmargin=*]
\item each \(C_t\) is a finite \(\mathbb J\)-stopping time; and
\item \(\calG_t\subseteq\mathcal J_{C_t}\) and \(\widehat P_t=Q_{C_t}\) for every \(t\geq0\).
\end{enumerate}
The public experiment is \textit{clock-dominated} if it has a disclosure clock \(C\) satisfying \(C_t\leq t\) for every \(t\geq0\) almost surely. A \textit{clock-dominated public policy} is a triple \((\G,C,\tau)\), where \(C\) is such a clock and \(\tau\) is an integrable public stopping rule.
\end{definition}
If \(C_t=s\), then \(\widehat P_t=Q_s\). Silence makes \(C\) flat, faster disclosure makes it steep, and releasing a stored interval makes it jump. Clock domination allows catch-up and jumps but requires \(C_t\leq t\) at every date. The requirements in \Cref{def:clock-domination} ensure that a public stopping rule \(\tau\) becomes a stopping time \(C_\tau\) for \(Q\). In \Cref{app:disclosureclock}, we construct disclosure clocks for a continuous public likelihood and delayed or batched release of a stored record.

\paragraph{The dynamics of garbling.} The possibility of garbling across time introduces a dynamic issue that is wholly absent from a static garbling problem. To see it, consider a discrete-time model with discrete signals. At any fixed date \(T\), the receiver's posterior after any reporting rule is a conditional expectation of the sender's date-\(T\) posterior and is, therefore, a mean-preserving contraction of \(P_T\) from the \textit{ex ante} perspective. 

Conditional on the sender's actual private history at \(T-1\), however, she faces a particular continuation experiment: the conditional distribution of \(P_T\) generated by the remaining signal. An intertemporal garbling--whether implemented through silence, delayed disclosure, or a reporting rule that pools information across dates--\textit{need not induce a mean-preserving contraction of this continuation distribution}. Because the receiver did not observe which date-\((T-1)\) history occurred, the reporting rule can pool terminal outcomes descended from different date-\((T-1)\) histories, including branches that the sender already knows did not occur. Consequently, \(\E[\widehat P_T\mid\calF_{T-1}^X]\) need not equal \(P_{T-1}\), and \(\widehat P_T\) may even take values outside the convex hull of the date-\(T\) posteriors that remain possible conditional on the sender's actual history.

Of course, the report is still a valid garbling of \(P_T\) from the \textit{ex ante} perspective, but it achieves this by recombining branches of the learning tree that are no longer available from the sender's current position. In the Brownian model, conditional support over every positive interval is too rich for this convex-hull criterion to be meaningful. The same issue instead appears as excessive cumulative speed: by date \(t\), intertemporal garbling can move the public posterior through more full-speed posterior time than has elapsed in calendar time. Clock domination precisely rules out this acceleration.

\subsection{Best Garbling Is No Garbling}\label{subsec:clock-bgng}

Let \(\F^P\) be the usual augmentation of the natural filtration of \(P\). Fix a bounded measurable terminal payoff \(F\colon(0,1)\to\R\) and an increasing convex delay cost \(c\colon[0,\infty)\to\R\). Define
\[
V^{\mathrm{garb}}_{F,c}
\coloneqq
\sup_{\substack{(\G,C,\tau)\text{ a clock-dominated}\\
\text{public policy}}}
\E[F(\widehat P_\tau)-c(\tau)]
\]
and
\[
V^0_{F,c}
\coloneqq
\sup_{\substack{\rho\text{ an }\F^P\text{-stopping time}\\
\E[\rho]<\infty}}
\E[F(P_\rho)-c(\rho)].
\]
Because \(F\) is bounded and \(c\) is bounded below, both values are finite.

\begin{theorem}\label{thm:clock-bgng}
Maintain \Cref{ass:binary-brownian-evidence}. Let \((\G,C,\tau)\) be a clock-dominated public policy, and set \(\nu\coloneqq\Law(\widehat P_\tau)\). There is an integrable \(\F^P\)-stopping time \(R^\nu\) such that \(\Law(P_{R^\nu})=\nu\), \(R^\nu\preceq_{\mathrm{cx}}C_\tau\), and \(C_\tau\leq\tau\) almost surely; and so \(R^\nu\icx\tau\).
\end{theorem}

The terminal posterior distributions agree. By \Cref{def:increasing-convex-order}, \(R^\nu\preceq_{\mathrm{cx}}C_\tau\) means that \(R^\nu\) and \(C_\tau\) have the same mean and that \(R^\nu\) is less dispersed. Clock domination implies \(C_\tau\leq\tau\); consequently, \(\E[F(P_{R^\nu})-c(R^\nu)]\geq\E[F(\widehat P_\tau)-c(\tau)]\).

Full transparency is clock-dominated with \(C_t=t\). Hence, \Cref{thm:clock-bgng} implies \(V^{\mathrm{garb}}_{F,c}=V^0_{F,c}\). The proof of \Cref{thm:clock-bgng} also produces:
\begin{enumerate}[label=(\roman*),nosep,noitemsep,leftmargin=*]
\item if \(\tau\leq T\) almost surely for some deterministic \(T<\infty\), then \(R^\nu\leq T\) almost surely;
\item if \(\E[c(\tau)]<\infty\), \(c\) is strictly increasing, and \(\Pp(C_\tau<\tau)>0\), then \(\E[c(R^\nu)]<\E[c(\tau)]\); and
\item if \(\E[c(C_\tau)]<\infty\), \(c\) is strictly convex, and \(\Law(R^\nu)\neq\Law(C_\tau)\), then \(\E[c(R^\nu)]<\E[c(C_\tau)]\).
\end{enumerate}
For the binary-action persuasion problem in \secref{sec:model}, take \(F(p)\coloneqq\1_{\{p\geq\bar P\}}\).

\paragraph{Proof sketch.} Fix a clock-dominated public policy \((\G,C,\tau)\), set \(\nu\coloneqq\Law(\widehat P_\tau)\), and retain the full-speed posterior \(Q\) associated with \(C\). Because \(C\) is a disclosure clock, \(U\coloneqq C_\tau\) is a stopping time for \(Q\) and \(Q_U=\widehat P_\tau\). Clock domination implies \(U\leq\tau\), so \(U\) is integrable and embeds \(\nu\) in \(Q\).

Root's timing comparison then provides a deterministic barrier \(\mathcal B^\nu\) whose first-entry time \(R_Q^\nu\coloneqq R_{\mathcal B^\nu}(Q)\) also embeds \(\nu\) and satisfies \(R_Q^\nu\preceq_{\mathrm{cx}}U\). Now use the same barrier for the posterior under full transparency, setting \(R^\nu\coloneqq R_{\mathcal B^\nu}(P)\). Since \(P\) and \(Q\) solve the same stochastic differential equation from the same initial condition,
\(\Law(R^\nu,P_{R^\nu}) = \Law(R_Q^\nu,Q_{R_Q^\nu})\). Hence, \(\Law(P_{R^\nu})=\nu\) and \(R^\nu\preceq_{\mathrm{cx}}C_\tau\). Combining this comparison with \(C_\tau\leq\tau\) produces \(R^\nu\icx\tau\) and so for every increasing convex delay cost \(c\),
\(\E[c(R^\nu)]
=
\E[c(R_Q^\nu)]
\leq
\E[c(C_\tau)]
\leq
\E[c(\tau)]\). The transparent rule preserves the terminal posterior distribution and lowers expected delay cost.

\subsection{Why the Clock Bound Matters}\label{subsec:clock-bound-necessity}

Without the clock bound, the sender can compress accumulated evidence into a coarse assessment and then communicate that assessment faster than the underlying evidence arrives. We now show that this possibility can make garbling strictly better than every transparent stopping rule. Specifically,

\begin{example}\label{ex:fast-binary-summary}
Maintain \Cref{ass:binary-brownian-evidence}. There are parameters, a finite, continuous, nondecreasing, convex delay cost, and an integrable public policy with continuous, strictly positive public likelihood such that the policy admits a disclosure clock, is not clock-dominated, and yields strictly greater payoff than every transparent stopping rule.
\end{example}

The construction is simple and laid out precisely in \Cref{app:exproof}. Set \(p_0=1/2\), and choose \(\kappa>0\) so that \(\Phi(\kappa/2)=2/3\), where \(\Phi\) is the standard normal distribution function. At date one, the sender retains only
\[
D\coloneqq\1_{\{L_1\geq0\}}
\qquad \text{and} \qquad
L_t\coloneqq\log\frac{P_t}{1-P_t}.
\]
The two realizations of \(D\) occur with equal probability and induce posteriors \(2/3\) and \(1/3\). We choose the persuasion threshold \(\bar P<2/3\) sufficiently close to \(2/3\) that the transparent posterior reaches \(\bar P\) by date one with probability strictly below \(1/2\).

The sender remains silent through date one. Thereafter, she releases noisy evidence about \(D\): the report has drift \(\lambda\) when \(D=1\), drift \(-\lambda\) when \(D=0\), and independent Brownian noise. Formally, for an independent Brownian motion \(V\),
\[
Y_t^\lambda
\coloneqq
\lambda(2D-1)(t-1)^++V_t-V_{t\wedge1}.
\]
She stops when the resulting public posterior first reaches either \(1-\bar P\) or \(\bar P\). By symmetry, the favorable boundary is reached with probability \(1/2\).

The first benefit of garbling comes from pooling. The summary \(D\) discards both the path and the magnitude of the evidence and retains only whether the date-one likelihood favors the high state. Its favorable realization pools moderately favorable histories with strongly favorable ones. Full transparency cannot pool these histories: it persuades only when each posterior path itself reaches \(\bar P\). Consequently, the summary supports persuasion on half of the histories even though transparency reaches the threshold by date one with strictly smaller probability.

The second benefit comes from the speed of release. Increasing \(\lambda\) does not change the terminal posteriors or the persuasion probability, but it shortens the communication period after date one at rate \(1/\lambda^2\). The public posterior is continuous and, thus, has a canonical disclosure clock. The terminal value of that clock has the distribution of a full-speed posterior's exit time from \((1-\bar P,\bar P)\), which does not shrink with \(\lambda\) and has an unbounded right tail. Calendar stopping time, in stark contrast, becomes concentrated arbitrarily close to date one. Accordingly, for sufficiently large \(\lambda\), the experiment admits a disclosure clock but cannot admit one satisfying \(C_t\leq t\).

A deadline-like convex delay cost makes this compression valuable. Choose a date \(T>1\) sufficiently close to one and take \(c(t)=M(t-T)^+\) with \(M\) large. Under transparency, persuasion by the relevant deadline occurs with probability strictly below \(1/2\), while later persuasion is offset by its delay cost. The garbled policy persuades with probability \(1/2\), and its expected post-date-one cost converges to zero as \(\lambda\) increases. Consequently, it strictly outperforms every transparent stopping rule for sufficiently large \(\lambda\).

The gain from garbling and the failure of the clock bound have the same source. Garbling first pools evidence histories into a useful one-bit summary; rapid release then compresses the posterior movement generated by that summary into less calendar time than clock domination permits.

\section{Comparative Statics}\label{sec:cs}

The preceding results reduce optimization within the clock-dominated class to transparent stopping. Let \(\F^P\) be the usual augmentation of the natural filtration of \(P\), so that transparent stopping rules are precisely the \(\F^P\)-stopping times, and define
\[H \coloneqq\inf\{t\geq0\colon P_t=\bar P\} \quad \text{and} \quad
\mathcal T_H \coloneqq\left\{\rho\colon \rho\text{ is an }\F^P\text{-stopping time},\ \rho\leq H\text{ a.s.},\ \E[\rho]<\infty\right\}.\]
For a cost \(c\), define
\[
J_c(\rho)\coloneqq\Pp(\rho=H)-\E[c(\rho)],
\quad
\mathcal V(c)\coloneqq\sup_{\rho\in\mathcal T_H}J_c(\rho),
\quad \text{and} \quad
\mathcal S(c)\coloneqq\argmax_{\rho\in\mathcal T_H}J_c(\rho).
\]
\begin{lemma}\label{lem:threshold-normalization}
Maintain \Cref{ass:binary-brownian-evidence}. Every integrable \(\F^P\)-stopping time \(\rho\) is dominated by \(\rho\wedge H\). If \(\rho\) is optimal for the transparent problem, then \(\rho\wedge H\) is also optimal and has the same persuasion probability.
\end{lemma}

\begin{lemma}\label{lem:transparent-optimal-stopping-lattice}
Maintain \Cref{ass:binary-brownian-evidence}. For every nonzero delay cost \(c\), the set \(\mathcal S(c)\) is nonempty, is closed under pairwise minimum and maximum, and contains a least and a greatest element under the order \(\rho\leq\eta\) almost surely.
\end{lemma}

The first lemma says that an optimal transparent rule never waits after the posterior has reached the persuasion threshold. The second says that optimal normalized rules have an ordered structure: taking the earlier or later of two optimal rules remains optimal, and there are well-defined earliest and latest optimal rules. Together, the lemmas imply that every element of \(\mathcal S(c)\) is optimal for the unrestricted transparent problem and that every transparent optimum has a representative in \(\mathcal S(c)\) with the same persuasion probability.

By \Cref{lem:transparent-optimal-stopping-lattice}, write \(\rho_c^-\) and \(\rho_c^+\) for the least and greatest elements of \(\mathcal S(c)\) under this order. Because \(\{\rho=H\}\subseteq\{\eta=H\}\) whenever \(\rho\leq\eta\leq H\), the minimum and maximum persuasion probabilities attained by optimal transparent stopping rules are \(\underline q(c)\coloneqq\Pp(\rho_c^-=H)\) and \(\overline q(c)\coloneqq\Pp(\rho_c^+=H)\), respectively.

\begin{definition}\label{def:stopping-strong-set-order}
For nonempty \(\mathcal A,\mathcal B\subseteq\mathcal T_H\), write \(\mathcal A\succeq_{\mathrm{SSO}}\mathcal B\) if, for every \(\rho\in\mathcal A\) and \(\eta\in\mathcal B\), \(\rho\vee\eta\in\mathcal A\) and \(\rho\wedge\eta\in\mathcal B\).
\end{definition}

\subsection{Delay Costs}\label{subsec:cost-comparative-statics}

\begin{definition}\label{def:higher-incremental-delay-cost}
A cost \(c_2\) has higher incremental delay costs than \(c_1\) if \(t\mapsto c_2(t)-c_1(t)\) is nondecreasing.
\end{definition}

\begin{proposition}\label{prop:higher-delay-cost-lower-persuasion}
Maintain \Cref{ass:binary-brownian-evidence}. If \(c_2\) has higher incremental delay costs than \(c_1\), then \(\mathcal S(c_1)\succeq_{\mathrm{SSO}}\mathcal S(c_2)\).
Consequently, \(\rho_{c_2}^-\leq\rho_{c_1}^-\), \(\rho_{c_2}^+\leq\rho_{c_1}^+\), \(\underline q(c_2)\leq\underline q(c_1)\), and \(\overline q(c_2)\leq\overline q(c_1)\)
\end{proposition}

Thus, making additional delay more costly shifts both the earliest and the latest optimal stopping rules earlier. Because reaching \(H\) is more likely under a later rule, both the minimum and maximum persuasion probabilities attained by optimal policies fall.

\subsection{Signal-to-Noise Ratio}\label{subsec:snr-comparative-statics}

To compare different signal-to-noise ratios on a common process, let \(P^1\) satisfy
\[
\dd P_u^1=P_u^1(1-P_u^1)\dd W_u,
\qquad \text{with }
P_0^1=p_0,
\]
and let \(\F^1=(\calF_u^1)_{u\geq0}\) be the usual augmentation of its natural filtration. Define \(H_1\coloneqq\inf\{u\geq0\colon P_u^1=\bar P\}\), and let \(\mathcal T_{H_1}^1\) be the \(\F^1\)-stopping times \(U\) satisfying \(U\leq H_1\) almost surely and \(\E[U]<\infty\). For each \(\kappa>0\), define
\[
d_\kappa(u)\coloneqq c\left(\frac{u}{\kappa^2}\right), \quad
J_\kappa^1(U)\coloneqq\Pp(U=H_1)-\E[d_\kappa(U)], \quad \text{and} \quad
\mathcal S^1(\kappa)\coloneqq\argmax_{U\in\mathcal T_{H_1}^1}J_\kappa^1(U).
\]
Applying \Cref{lem:transparent-optimal-stopping-lattice} to this common process produces least and greatest elements \(U_\kappa^-\) and \(U_\kappa^+\). Define \(\underline q_{\mathrm{snr}}(\kappa)\coloneqq\Pp(U_\kappa^-=H_1)\) and \(\overline q_{\mathrm{snr}}(\kappa)\coloneqq\Pp(U_\kappa^+=H_1)\).

For each \(\kappa>0\), set \(P_t^\kappa\coloneqq P_{\kappa^2t}^1\). Then \(W_t^\kappa\coloneqq\kappa^{-1}W_{\kappa^2t}\) is a Brownian motion and
\[
\dd P_t^\kappa=\kappa P_t^\kappa(1-P_t^\kappa)\dd W_t^\kappa.
\]
Moreover, the usual natural filtration of \(P^\kappa\) at date \(t\) is \(\calF_{\kappa^2t}^1\), so every signal-to-noise ratio is a deterministic time change of the same posterior process. The map \(\rho\mapsto U=\kappa^2\rho\) is a bijection between integrable stopping times for \(P^\kappa\) and integrable stopping times for \(P^1\). It maps the first hitting time of \(\bar P\) under \(P^\kappa\) to \(H_1\), preserves the persuasion event, and transforms calendar cost \(c(\rho)\) into information-time cost \(d_\kappa(U)\). A higher \(\kappa\) makes each unit of information time require less calendar time.
\begin{proposition}\label{prop:higher-snr-higher-persuasion}
If \(\kappa_2\geq\kappa_1\), then \(\mathcal S^1(\kappa_2)\succeq_{\mathrm{SSO}}\mathcal S^1(\kappa_1)\). Consequently, \(U_{\kappa_1}^-\leq U_{\kappa_2}^-\), \(U_{\kappa_1}^+\leq U_{\kappa_2}^+\), \(\underline q_{\mathrm{snr}}(\kappa_1)\leq\underline q_{\mathrm{snr}}(\kappa_2)\), \(\overline q_{\mathrm{snr}}(\kappa_1)\leq\overline q_{\mathrm{snr}}(\kappa_2)\).\end{proposition}

Accordingly, a higher signal-to-noise ratio shifts both extremal optimal rules later in information time. Waiting for more information is cheaper in calendar time, so both extremal persuasion probabilities rise.

\section{Receiver Stopping}\label{sec:uncommitted-receiver}

In the preceding sections, the receiver has no waiting cost and therefore no incentive to act before the sender stops providing information. We now allow her to stop and act earlier. The sender remains committed to her disclosure policy, including when it ceases to reveal information. Retain the binary terminal action from \secref{sec:model}. At posterior \(p\), action \(a\in\{0,1\}\) delivers the receiver expected payoff \(u_R(p,a)\), which is affine in \(p\), and action one is optimal if and only if \(p\geq\bar P\). Define \(r(p)\coloneqq\max_{a\in\{0,1\}}u_R(p,a)\), which is bounded, continuous, and convex. The receiver incurs the linear waiting cost \(\kappa_Rt\), where \(\kappa_R>0\).

The sender commits to a pair
\(\mathsf S=(\mathbb G^{\mathsf S},C)\), where \(\mathbb G^{\mathsf S}\) is a public experiment and \(C\) is a disclosure clock satisfying \(C_t\leq t\) for every \(t\geq0\) almost surely. Write \(\widehat P^{\mathsf S}\) for the associated public posterior.
\begin{definition}\label{def:receiver-equilibrium}
A \textit{receiver equilibrium} is a pair \((\mathsf S,\rho)\) such that \(\rho\) is a \(\mathbb G^{\mathsf S}\)-stopping time with \(\E[\rho]<\infty\) and \(\E\left[r(\widehat P_\rho^{\mathsf S})-\kappa_R\rho\right] \geq \E\left[r(\widehat P_\tau^{\mathsf S})-\kappa_R\tau\right]\) for every \(\mathbb G^{\mathsf S}\)-stopping time \(\tau\) with \(\E[\tau]<\infty\).
\end{definition}

\begin{definition}\label{def:unbanked-clock}
A disclosure clock \(C\) is \textit{unbanked} if, almost surely, \(C_t-C_s\leq t-s\) for every \(0\leq s\leq t\).
\end{definition}

An unbanked clock is continuous and satisfies \(C_t\leq t\). It permits silence but not later disclosure above full speed. This stronger restriction matters because the receiver may stop before the experiment ends.

Fix a receiver equilibrium \((\mathsf S,\rho)\) with an unbanked disclosure clock \(C\). Because \(C\) is a disclosure clock, there exist a usual filtration \(\mathbb J\) and a \(\mathbb J\)-adapted full-speed posterior \(Q\) satisfying the requirements in \Cref{def:clock-domination}. Fix such \(\mathbb J\) and \(Q\), and set \(U\coloneqq C_\rho\). Then \(U\) is a \(\mathbb J\)-stopping time and \(\widehat P_\rho^{\mathsf S}=Q_U\).

We first relate stopping \(Q\) before \(U\) to stopping the original experiment in calendar time. For \(u\geq0\), define \(T_u^+\coloneqq\inf\{t\geq0\colon C_t>u\}\) and \(\theta_u\coloneqq\rho\wedge T_u^+\), where \(\inf\varnothing=\infty\). The date \(\theta_u\) is the earlier of receiver stopping and the calendar date at which the disclosure clock first moves strictly above \(u\). The family \((\theta_u)_{u\geq0}\) is increasing and right-continuous.

Define \(\calH_u^0\coloneqq\calG_{\theta_u}^{\mathsf S}\), and write \(\mathbb H^0=(\calH_u^0)_{u\geq0}\). If \(u_n\downarrow u\), then \(\theta_{u_n}\downarrow\theta_u\), and right-continuity of \(\mathbb G^{\mathsf S}\) implies \(\bigcap_n\calH_{u_n}^0=\calH_u^0\). Hence, \(\mathbb H^0\) is right-continuous. Let \(\mathbb H=(\calH_u)_{u\geq0}\) be its completion. The strict inequality in the definition of \(T_u^+\) selects the end of a flat stretch of the clock. This retains any state-independent public information revealed while the clock is paused.

\begin{lemma}\label{lem:receiver-operational-pullback}
Maintain \Cref{ass:binary-brownian-evidence}. Let \((\mathsf S,\rho)\) be a receiver equilibrium whose fixed clock \(C\) is unbanked. Then \(U\) is an \(\mathbb H\)-stopping time. For every \(u\geq0\), \(C_{\theta_u}=u\wedge U\) and \(\widehat P_{\theta_u}^{\mathsf S}=Q_{u\wedge U}\). If \(V\leq U\) is an \(\mathbb H\)-stopping time, then \(\theta_V\) is a \(\mathbb G^{\mathsf S}\)-stopping time, \(\widehat P_{\theta_V}^{\mathsf S}=Q_V\), and \(\theta_V-V\leq\rho-U\) almost surely.
\end{lemma}

If the receiver would stop \(Q\) at \(V\leq U\), she can instead stop the original experiment at \(\theta_V\). The two rules produce the same posterior. Moreover, \(\theta_V-V\leq\rho-U\), so the alternative rule adds no more calendar delay than the equilibrium rule itself adds.

\begin{lemma}\label{lem:receiver-stopped-experiment-transplant}
Maintain \Cref{ass:binary-brownian-evidence}. There is a stopping time \(\sigma_D\) for the filtration generated by \(P\) and independent date-zero randomization such that
\[\left(\left(P_{u\wedge\sigma_D}\right)_{u\geq0},\left(\1_{\{\sigma_D\leq u\}}\right)_{u\geq0}\right)\stackrel{d}{=}\left(\left(Q_{u\wedge U}\right)_{u\geq0},\left(\1_{\{U\leq u\}}\right)_{u\geq0}\right).\] Moreover, every stopping time \(\eta\leq\sigma_D\) for the usual natural filtration of the process on the left has a stopping time \(V\leq U\) for the usual natural filtration of the process on the right such that \((P_\eta,\eta)\stackrel{d}{=}(Q_V,V)\).
\end{lemma}
The lemma constructs \(\sigma_D\) so that the stopped experiment under \(P\) has the same law as the stopped experiment under \(Q\) at \(U\). Moreover, each stopping rule \(\eta\leq\sigma_D\) under transparency has a stopping rule \(V\leq U\) satisfying \((P_\eta,\eta)\stackrel{d}{=}(Q_V,V)\). By \Cref{lem:receiver-operational-pullback}, \(\theta_V\) is a stopping rule under \(\mathsf S\), produces posterior \(Q_V\), and satisfies \(\theta_V-V\leq\rho-U\).

\begin{definition}\label{def:transparent-acceleration}
The policy \(D\), called the \textit{transparent acceleration}, reveals \(P\), the posterior under full transparency, until \(\sigma_D\), publicly announces termination at \(\sigma_D\), and reveals no further information afterward. Its public filtration \(\mathbb G^D=(\calG_t^D)_{t\geq0}\) is the usual augmentation of \(\mathbb G^{D,0}=(\calG_t^{D,0})_{t\geq0}\), where
\(\calG_t^{D,0}\coloneqq\sigma\left(P_{s\wedge\sigma_D},\1_{\{\sigma_D\leq s\}}\colon 0\leq s\leq t\right)\).
\end{definition}

The transparent acceleration has disclosure clock \(C_t^D=t\wedge\sigma_D\), i.e., it reveals at full speed until \(\sigma_D\) and then stops permanently. The verification is included in the proof of \Cref{thm:uncommitted-receiver-acceleration}.

\begin{theorem}\label{thm:uncommitted-receiver-acceleration}
Maintain \Cref{ass:binary-brownian-evidence}. Let \((\mathsf S,\rho)\) be a receiver equilibrium with unbanked disclosure clock \(C\). Then \(\sigma_D\) is a receiver best response to \(D\). The sender's payoff satisfies
\(\E\left[\1_{\{P_{\sigma_D}\geq\bar P\}}-c(\sigma_D)\right]
\geq
\E\left[\1_{\{\widehat P_\rho^{\mathsf S}\geq\bar P\}}-c(\rho)\right]\), and the receiver's payoff satisfies
\(\E[r(P_{\sigma_D})-\kappa_R\sigma_D]
=
\E[r(\widehat P_\rho^{\mathsf S})-\kappa_R\rho]
+\kappa_R\E[\rho-C_\rho]\).
\end{theorem}

By \Cref{lem:receiver-stopped-experiment-transplant}, \(P_{\sigma_D}\) and \(\widehat P_\rho^{\mathsf S}\) have the same distribution, while \(\sigma_D\) and \(C_\rho\) have the same distribution. Since \(C_\rho\leq\rho\), the sender's terminal payoff is unchanged and her expected delay cost falls. The receiver's terminal value is also unchanged, while her expected waiting cost falls by exactly \(\kappa_R\E[\rho-C_\rho]\).

\begin{corollary}\label{cor:uncommitted-receiver-bgng}
Maintain \Cref{ass:binary-brownian-evidence}. Every receiver equilibrium whose fixed clock is unbanked has a transparent acceleration that remains optimal for the receiver and benefits both players.
\end{corollary}

\bibliography{embedbib}

\appendix

\section{Microfounding the Reporting Environment}\label[appendix]{app:microfound}

We provide a concrete microfoundation for \Cref{ass:binary-brownian-evidence}. Let \((\Omega^r,\mathcal R,\Q)\) carry a random seed \(\xi\),\footnote{An initial random variable \(\xi\) used for date-zero randomization.} an \(m\)-dimensional Brownian motion \(V\), and a Poisson random measure \(\Pi\) on \(\R_+\times E\) with deterministic compensator \(\dd t\varrho(\dd z)\), where \(E\) is a standard Borel space and \(\varrho\) is \(\sigma\)-finite. Assume that \(\xi\), \(V\), and \(\Pi\) are mutually independent under \(\Q\), and let \(\mathbb U=(\mathcal U_t)_{t\geq0}\) be the \(\Q\)-usual augmentation of the filtration they generate, with \(\xi\in\mathcal U_0\). Under state \(i\in\{h,\ell\}\), take \(\Pp^i\coloneqq\Pp_X^i\otimes\Q\), where \(\Pp_X^i\) is the state-\(i\) law of \(X\). Thus, only the law of the evidence coordinate varies with the state.

Let \(\F^{r,0}\coloneqq(\calF_t^{r,0})_{t\geq0}\), where \(\calF_t^{r,0}\coloneqq\calF_t^X\vee\mathcal U_t\), and let \(\F^r=(\calF_t^r)_{t\geq0}\) be its usual augmentation under \(\Pp\). The filtration \(\F^r\) records the evidence and the state-independent randomization available to the reporting device through each date.

The sender may commit to any \(d\)-dimensional c\`adl\`ag \(\F^r\)-adapted report \(Y\) that admits the representation
\[\tag{A1}\label{eq:primitive-report}
Y_t
=
Y_0+D_t^Y
+\int_0^t\beta_s \dd X_s
+\int_0^t\gamma_s \dd V_s
+\int_{(0,t]\times E}\delta_s(z) \widetilde\Pi(\dd s,\dd z),\]
where \(\widetilde\Pi(\dd s,\dd z)\coloneqq\Pi(\dd s,\dd z)-\dd s \varrho(\dd z)\). Here, \(Y_0\) is \(\calF_0^r\)-measurable, \(D_0^Y=0\), and \(D^Y\) is an \(\F^r\)-adapted c\`adl\`ag process with finite variation on every bounded interval. The integrands \(\beta\), \(\gamma\), and \(\delta\) are \(\F^r\)-adapted and make the three stochastic integrals well defined on every bounded interval under either state. The adapted coefficients may depend on earlier evidence, reports, and auxiliary draws, while \(D^Y\) may also jump in response to information available at the jump date. Let \(\G^Y\) denote the usual augmentation of the natural filtration of \(Y\).

\paragraph{Report taxonomy.} \eqref{eq:primitive-report} contains the following reporting technologies.
\begin{enumerate}[label=(\roman*),noitemsep,nosep,leftmargin=*]
\item \emph{Full transparency and continuous reports.}
Taking \(\beta_t=1\) and setting the remaining terms equal to zero is full transparency. With no Poisson term and continuous \(D^Y\), the same equation permits multidimensional continuous reports that transform the evidence according to the earlier history and may add independent Gaussian noise.

\item \emph{Marked and categorical messages.}
A finite or countable subset of \(E\) may index message labels. The reporting rule may make the arrival rate and the distribution of labels depend on earlier evidence, so both the arrival of a message and its continued absence may be informative.

\item \emph{Silence, delayed disclosure, and batches.}
If the committed rule makes \(D^Y\) constant and \(\beta\), \(\gamma\), and \(\delta\) vanish on an interval, the report is silent there. If the duration of silence depends on earlier evidence, continued silence may itself be informative. Jumps of \(D^Y\) or of the marked component permit delayed announcements, summaries of earlier evidence, and batched disclosure.

\item \emph{L\'evy-type reports.}
The auxiliary Brownian and Poisson drivers permit  continuous
and jump noise.  Time-homogeneous characteristics produce ordinary L\'evy
components, while predictable characteristics produce more general It\^o-L\'evy
reports and random announcement times.

\item \emph{Hybrid reports.}
A single multidimensional report may combine continuous noisy disclosure,
categorical alerts, silence, random report dates, and delayed or batched
announcements.
\end{enumerate}

\begin{proposition}
\label{prop:primitive-reports-preserve-evidence}
In the product construction above, let \(\widetilde{\F}=(\widetilde{\calF}_t)_{t\geq0}\) be any usual filtration satisfying \(\calF_t^X\subseteq\widetilde{\calF}_t\subseteq\calF_t^r\) for every \(t\geq0\). Then \(\widetilde{\F}\) satisfies \Cref{ass:binary-brownian-evidence}. In particular, the conclusion holds for the usual augmentation of \((\calF_t^X\vee\calG_t^Y)_{t\geq0}\).
\end{proposition}
\begin{proof}[Proof of \Cref{prop:primitive-reports-preserve-evidence}]
We first verify the Brownian-motion requirement. Under \(\Pp^\ell\), the process \(B_t^\ell=(X_t-\mu_\ell t)/\sigma\) is Brownian in \(\F^X\). The product construction makes \(\mathbb U\) independent of the entire evidence process. Hence, for \(0\leq s<t\), the increment \(B_t^\ell-B_s^\ell\) is independent of \(\calF_s^{r,0}=\calF_s^X\vee\mathcal U_s\) and is Gaussian with mean zero and variance \(t-s\). Thus, \(B^\ell\) is Brownian in \(\F^{r,0}\). After the usual augmentation, \(B^\ell\) remains a continuous martingale with quadratic variation \(t\), so L\'evy's characterization implies that it is an \(\F^r\)-Brownian motion. Since \(\F^X\subseteq\widetilde{\F}\subseteq\F^r\), it is also a \(\widetilde{\F}\)-Brownian motion.

We next verify the likelihood-ratio formula. Let \(A_X\in\calF_t^X\) and \(A_U\in\mathcal U_t\). The product structure implies \(\Pp^h(A_X\cap A_U)=\Pp_X^h(A_X)\Q(A_U)=\E^\ell[Z_t\1_{A_X}]\Q(A_U)=\E^\ell[Z_t\1_{A_X\cap A_U}]\). The events \(A_X\cap A_U\) form a \(\pi\)-system that generates \(\calF_t^{r,0}\). The monotone-class theorem, thus, implies that \(\left.\frac{\dd\Pp^h}{\dd\Pp^\ell}\right|_{\calF_t^{r,0}}=Z_t\). If \(A\in\bigcap_{u>t}\calF_u^{r,0}\), then \(\Pp^h(A)=\E^\ell[Z_u\1_A]\) for every \(u>t\). Since \(Z_u\to Z_t\) in \(L^1(\Pp^\ell)\) as \(u\downarrow t\), letting \(u\downarrow t\) yields \(\Pp^h(A)=\E^\ell[Z_t\1_A]\). Because \(\Pp=p_0\Pp^h+(1-p_0)\Pp^\ell\) and both prior probabilities are positive, every \(\Pp\)-null event is null under both state-contingent laws. The identity, therefore, extends to the completion, so \(\left.\frac{\dd\Pp^h}{\dd\Pp^\ell}\right|_{\calF_t^r}=Z_t\).

Finally, \(Z_t\) is \(\calF_t^X\)-measurable, hence, \(\widetilde{\calF}_t\)-measurable. Restricting the preceding formula to \(\widetilde{\calF}_t\) yields \(\left.\frac{\dd\Pp^h}{\dd\Pp^\ell}\right|_{\widetilde{\calF}_t}=Z_t\).
This proves both parts of \Cref{ass:binary-brownian-evidence}.
\end{proof}

The product construction limits what the reporting device may use: all state information enters through \(X\), while \(\xi\), \(V\), and \(\Pi\) only randomize how the device transforms and releases that evidence.

\section{Disclosure Clock Representations}\label[appendix]{app:disclosureclock}

\subsection{Continuous Disclosure and the Canonical Clock}\label[appendix]{app:proof-inverse-clock}

Suppose the public likelihood \(\Lambda\) is continuous and strictly positive. Define \(N_t\coloneqq\int_0^t\Lambda_s^{-1}\dd\Lambda_s\) and \(I_t\coloneqq\langle N\rangle_t/\kappa^2\). Under full transparency, \(I_t=t\).

\begin{proposition}\label{prop:inverse-clock}
Maintain \Cref{ass:binary-brownian-evidence}. If the public likelihood is continuous and strictly positive, then \(I\) is a disclosure clock in the sense of \Cref{def:clock-domination}.
\end{proposition}

\begin{proof}[Proof of \Cref{prop:inverse-clock}]
We proceed through four claims.

\begin{claim}\label{claim:posterior-information-quadratic-variation}
The normalized martingale part of the public posterior has quadratic variation \(I\).
\end{claim}

Write \(f(x)\coloneqq p_0x/(1-p_0+p_0x)\), so \(\widehat P_t=f(\Lambda_t)\). Since \(xf'(x)=f(x)(1-f(x))\), It\^o's formula for quadratic variation yields
\[
\dd\langle\widehat P\rangle_t=\widehat P_t^2(1-\widehat P_t)^2\dd\langle N\rangle_t=\kappa^2\widehat P_t^2(1-\widehat P_t)^2\dd I_t.
\]
Under \(\Pp\), the process \(\widehat P\) is a bounded continuous \(\G\)-martingale. Define
\[
M_t\coloneqq\int_0^t\frac{1}{\kappa\widehat P_s(1-\widehat P_s)}\dd\widehat P_s.
\]
The public posterior remains in \((0,1)\) at every finite time because \(\Lambda\) is strictly positive and finite. The integral is well defined by localization, and by the preceding quadratic-variation identity, \(\langle M\rangle_t=I_t\).

\begin{claim}\label{claim:information-time-brownian-motion}
On an enlarged probability space, there is a Brownian motion \(W\) such that \(M_t=W_{I_t}\) for every calendar date \(t\).
\end{claim}

Set \(\zeta\coloneqq I_\infty\) and, for \(u\geq0\), define \(T_u\coloneqq\inf\{t\geq0\colon I_t>u\}\).
Each \(T_u\) is a \(\G\)-stopping time. On \(\{\zeta<\infty\}\), the continuous local-martingale convergence theorem delivers a finite limit \(M_\infty\coloneqq\lim_{t\to\infty}M_t\). Define \(\beta_u\coloneqq M_{T_u}\), with the convention \(M_{T_u}=M_\infty\) when \(T_u=\infty\), and let \(\mathbb K=(\mathcal K_u)_{u\geq0}\) be the usual augmentation of the time-changed filtration \(\mathcal K_u\coloneqq\calG_{T_u}\). The continuous local-martingale time-change theorem implies that \(\beta\) is a continuous \(\mathbb K\)-local martingale, that \(\zeta\) is a \(\mathbb K\)-stopping time, and that \(\langle\beta\rangle_u=u\wedge\zeta\).

Consider an extension of the probability space carrying a Brownian motion \(\widetilde W\) independent of \(\mathcal K_\infty\), and let \(\mathbb H=(\calH_u)_{u\geq0}\) be the usual augmentation of \((\mathcal K_u\vee\sigma(\widetilde W_s\colon 0\leq s\leq u))_{u\geq0}\). Define \(W_u\coloneqq\beta_u+\int_0^u\1_{\{s>\zeta\}}\dd\widetilde W_s\).
The integrand is predictable because \(\zeta\) is a \(\mathbb K\)-stopping time. The process \(W\) is a continuous \(\mathbb H\)-local martingale. Independence implies zero cross-variation between its two terms, so
\[
\langle W\rangle_u=u\wedge\zeta+\int_0^u\1_{\{s>\zeta\}} \dd s=u.
\]
Thus, L\'evy's characterization theorem implies that \(W\) is an \(\mathbb H\)-Brownian motion. By construction, \(\calG_{T_u}\subseteq\calH_u\) for every \(u\geq0\).

For every \(t\geq0\), \(T_{I_t}\geq t\), and \(I\) is constant on \([t,T_{I_t}]\). Since a continuous local martingale is constant on every interval on which its quadratic variation is constant, \(\beta_{I_t}=M_{T_{I_t}}=M_t\).
The independent continuation does not move before \(\zeta\), and, hence, \(M_t=W_{I_t}\) for every \(t\geq0\).

\begin{claim}\label{claim:public-posterior-full-speed-representation}
Each \(I_t\) is a stopping time in the information-time filtration, and \(\widehat P_t=Q_{I_t}\) for a full-speed posterior \(Q\).
\end{claim}

For every \(t\geq0\), \(\{I_t\leq u\}=\{t\leq T_u\}\in\calG_{T_u}\subseteq\calH_u\) means that \(I_t\) is an \(\mathbb H\)-stopping time. If \(A\in\calG_t\), then \(A\cap\{I_t\leq u\}=A\cap\{t\leq T_u\}\in\calG_{T_u}\subseteq\calH_u\), so \(A\in\calH_{I_t}\). Thus, \(\calG_t\subseteq\calH_{I_t}\). Let \(Q\) be the unique \([0,1]\)-valued strong solution to
\[
\dd Q_u=\kappa Q_u(1-Q_u)\dd W_u,\qquad \text{with } Q_0=p_0.
\]
The stochastic time-change theorem implies \(Q_{I_t}=p_0+\int_0^t\kappa Q_{I_s}(1-Q_{I_s})\dd M_s\), while \(\widehat P_t=p_0+\int_0^t\kappa\widehat P_s(1-\widehat P_s)\dd M_s\) by the definition of \(M\).
The coefficient is Lipschitz on \([0,1]\), so pathwise uniqueness yields \(\widehat P_t=Q_{I_t}\) for every \(t\geq0\).

\begin{claim}\label{claim:stopped-information-time}
For every public stopping rule \(\tau\), the information time \(U\coloneqq I_\tau\) is a stopping time, \(Q_U=\widehat P_\tau\), and the stopped posterior is uniformly integrable.
\end{claim}

Continuity and monotonicity of \(I\) imply \(I_s\leq u\) if and only if \(s\leq T_u\). Consequently, for \(U\coloneqq I_\tau\), \(\{U\leq u\}=\{\tau\leq T_u\}\in\calG_{T_u}\subseteq\calH_u\).
Thus \(U\) is an \(\mathbb H\)-stopping time. The process identity produces \(Q_U=\widehat P_\tau\). Finally, \(Q\in[0,1]\), so \((Q_{u\wedge U})_{u\geq0}\) is bounded and, consequently, uniformly integrable.
\end{proof}

\subsection{Stored Disclosure and Discontinuous Clocks}\label[appendix]{app:stored-disclosure}

We now provide a construction with delayed disclosure that admits a disclosure clock. The underlying posterior is continuous in evidence time, while its release in calendar time may be delayed or batched. Let \(\mathbb K=(\mathcal K_u)_{u\geq0}\) be a filtration, let \(A=(A_t)_{t\geq0}\) be a nonnegative process, and write
\(\check P_u\coloneqq\Pp(\mu=\mu_h\mid\mathcal K_u)\).

\begin{definition}\label{def:stored-disclosure-experiment} The pair \((\mathbb K,A)\) is a \textit{stored-disclosure experiment} if \(\mathbb K\) is a usual subfiltration of \(\F\), \(\check P\) has continuous paths in \((0,1)\), \(A\) is increasing and right-continuous with \(A_0=0\), and each \(A_t\) is a finite \(\mathbb K\)-stopping time satisfying \(A_t\leq t\).
\end{definition}

For a stored-disclosure experiment, let \(\mathbb G^A=(\calG_t^A)_{t\geq0}\) be the usual augmentation of
\(\calG_t^{A,0}
\coloneqq
\sigma(A_s\colon0\leq s\leq t)\vee\mathcal K_{A_t}\), and write
\(\widehat P_t^A
\coloneqq
\Pp(\mu=\mu_h\mid\calG_t^A)\). Finally, set
\[
\Gamma_u
\coloneqq
\int_0^u
\frac{\dd\langle\check P\rangle_s}
{\kappa^2\check P_s^2(1-\check P_s)^2},
\qquad \text{with }
C_t
\coloneqq
\Gamma_{A_t}.
\]

The process \(A_t\) is the latest date of the underlying record released by calendar date \(t\). A jump from \(a\) to \(b\) releases the record generated over \((a,b]\), rather than a summary constructed at the release date. Accordingly, the disclosure clock \(C_t=\Gamma_{A_t}\) may jump even though \(\Gamma\) is continuous.

\begin{proposition}\label{prop:stored-disclosure-representation}
Maintain \Cref{ass:binary-brownian-evidence}. Every stored-disclosure experiment admits a disclosure clock \(C_t=\Gamma_{A_t}\).
\end{proposition}

\begin{proof}
Because \(\check P_u=\E[\1_{\{\mu=\mu_h\}}\mid\mathcal K_u]\) and \(A_t\) is a finite \(\mathbb K\)-stopping time, optional sampling at the stopped sigma-field delivers \(\E[\1_{\{\mu=\mu_h\}}\mid\mathcal K_{A_t}]=\check P_{A_t}\). If \(s\leq t\), then \(A_s\leq A_t\), so \(\mathcal K_{A_s}\subseteq\mathcal K_{A_t}\). Since \(A_s\) is \(\mathcal K_{A_s}\)-measurable, the release history through date \(t\) belongs to \(\mathcal K_{A_t}\). Right-continuity and completion preserve this inclusion, and hence \(\calG_t^A\subseteq\mathcal K_{A_t}\). The posterior \(\check P_{A_t}\) is public at date \(t\), so the tower property yields \(\widehat P_t^A=\check P_{A_t}\).

Define \(M_u\coloneqq\int_0^u\frac{1}{\kappa\check P_s(1-\check P_s)}\dd\check P_s\). Then \(M\) is a continuous \(\mathbb K\)-local martingale satisfying \(\langle M\rangle_u=\Gamma_u\). For \(v\geq0\), let \(T_v^\Gamma\coloneqq\inf\{u\geq0\colon\Gamma_u>v\}\). The continuous local-martingale time-change theorem, followed by an independent Brownian continuation beyond \(\Gamma_\infty\), produces a usual filtration \(\mathbb J\), a \(\mathbb J\)-Brownian motion, and a full-speed posterior \(Q\) such that \(\check P_u=Q_{\Gamma_u}\) and \(\mathcal K_{T_v^\Gamma}\subseteq\mathcal J_v\).

Continuity and monotonicity of \(\Gamma\) imply \(\{C_t\leq v\}=\{A_t\leq T_v^\Gamma\}\). Since \(A_t\) and \(T_v^\Gamma\) are \(\mathbb K\)-stopping times, the event on the right belongs to \(\mathcal K_{T_v^\Gamma}\subseteq\mathcal J_v\), so each \(C_t\) is a \(\mathbb J\)-stopping time. If \(B\in\calG_t^A\), then, for every \(v\geq0\),
\[
B\cap\{C_t\leq v\}=B\cap\{A_t\leq T_v^\Gamma\}\in\mathcal K_{T_v^\Gamma}\subseteq\mathcal J_v.
\]
Thus, \(\calG_t^A\subseteq\mathcal J_{C_t}\), and \(\widehat P_t^A=\check P_{A_t}=Q_{C_t}\).

Finally, if \(\tau\) is a public stopping rule, right-continuity and monotonicity of \(C\) produce, for every \(v>0\), \(\{C_\tau<v\}=\bigcup_{r\in\mathbb Q_+}\left(\{\tau<r\}\cap\{C_r<v\}\right)\).
The inclusion \(\calG_r^A\subseteq\mathcal J_{C_r}\) makes every event in the union \(\mathcal J_v\)-measurable. As a result, right-continuity of \(\mathbb J\) makes \(C_\tau\) a \(\mathbb J\)-stopping time.
\end{proof}

\section{Proofs and Derivations}\label[appendix]{app:bgng-proofs}

\subsection{Proof of \texorpdfstring{\Cref{prop:mean-time-efficiency}}{the proposition}}\label[appendix]{app:proof-mean-time-efficiency}

\begin{proof}[Proof of \Cref{prop:mean-time-efficiency}]
We proceed through two claims.

\begin{claim}\label{claim:public-mean-time-lower-bound}
Every public stopping rule satisfies the stated lower bound on expected calendar time.
\end{claim}

Let \(\Pp_{\tau,\G}^i\) denote the restriction of \(\Pp^i\) to \(\calG_\tau\). Via Bayes' rule,
\[
\log\frac{\dd\Pp_{\tau,\G}^h}{\dd\Pp_{\tau,\G}^\ell}
=
\log\frac{\widehat P_\tau}{1-\widehat P_\tau}
-\log\frac{p_0}{1-p_0}.
\]
Consequently, using \(\E[\widehat P_\tau]=p_0\),
\[
\begin{aligned}
&p_0\mathrm D(\Pp_{\tau,\G}^h\Vert\Pp_{\tau,\G}^\ell)
+(1-p_0)\mathrm D(\Pp_{\tau,\G}^\ell\Vert\Pp_{\tau,\G}^h)\\
&=\E\left[(2\widehat P_\tau-1)\left(\log\frac{\widehat P_\tau}{1-\widehat P_\tau}-\log\frac{p_0}{1-p_0}\right)\right]=\E[\Psi(\widehat P_\tau)]-\Psi(p_0).
\end{aligned}
\]

Let \(\Pp_{\tau,\F}^i\) denote the restriction of \(\Pp^i\) to \(\calF_\tau\). The stopped full likelihood ratio is \(Z_\tau\). To verify this claim, fix \(A\in\calF_\tau\). For every \(n\), \(A\cap\{\tau\leq n\}\in\calF_n\), and bounded-time optional sampling delivers \(\Pp^h(A\cap\{\tau\leq n\})=\E^\ell[Z_n\1_{A\cap\{\tau\leq n\}}]=\E^\ell[Z_\tau\1_{A\cap\{\tau\leq n\}}]\).
Letting \(n\to\infty\) yields \(\left.\frac{\dd\Pp^h}{\dd\Pp^\ell}\right|_{\calF_\tau}=Z_\tau\). Under \(\Pp^h\), \(\log Z_\tau=\kappa B_\tau^h+\frac{\kappa^2}{2}\tau\).
Since \(\E^h[\tau]<\infty\), the stopped Brownian motion has zero mean, and so \(\mathrm D(\Pp_{\tau,\F}^h\Vert\Pp_{\tau,\F}^\ell)=\frac{\kappa^2}{2}\E^h[\tau]\). Similarly, \(\mathrm D(\Pp_{\tau,\F}^\ell\Vert\Pp_{\tau,\F}^h)=\frac{\kappa^2}{2}\E^\ell[\tau]\).
The data-processing inequality for relative entropy and \(\calG_\tau\subseteq\calF_\tau\) produce
\[\E[\Psi(\widehat P_\tau)]-\Psi(p_0) \leq\frac{\kappa^2}{2}\left(p_0\E^h[\tau]+(1-p_0)\E^\ell[\tau]\right) =\frac{\kappa^2}{2}\E[\tau].
\]

\begin{claim}\label{claim:full-speed-mean-time-identity}
For the full-speed posterior, expected stopping time is determined by the terminal posterior distribution through the stated identity.
\end{claim}

For the full-speed posterior \(Q\), direct differentiation produces \(\Psi''(p)=\frac{1}{p^2(1-p)^2}\).
Applying It\^o's formula after localization yields
\[
\dd\Psi(Q_t)=\frac{\kappa^2}{2} \dd t+\kappa Q_t(1-Q_t)\Psi'(Q_t) \dd W_t.
\]
The function \(p\mapsto p(1-p)\Psi'(p)=2p(1-p)\log\frac{p}{1-p}+2p-1\) is bounded on \((0,1)\). Since \(\E[S]<\infty\), the stopped stochastic integral is square integrable and has mean zero, and so passing through the localization, \(\E[\Psi(Q_S)]-\Psi(p_0)=\frac{\kappa^2}{2}\E[S]\).\end{proof}

\subsection{Proof of \texorpdfstring{\Cref{thm:clock-bgng}}{the theorem}}\label[appendix]{app:proof-clock-bgng}

First, we recall Root's timing comparison.

\begin{lemma}\label{lem:root-icx}
Suppose \(\eta\) is locally Lipschitz, strictly positive in the interior of \(\mathcal I\), and satisfies \(\eta(x)^2\leq C(1+x^2)\). Suppose the endpoints are either absorbing or inaccessible and that every finite inaccessible endpoint is approached with positive probability as time tends to infinity. Let \(\mu_0\) and \(\mu_1\) be centered integrable probability measures on \(\overline{\mathcal I}\) satisfying \(\mu_0\preceq_{\mathrm{cx}}\mu_1\). Let \(R\) be the Root stopping time that embeds \(\mu_1\) from \(\mu_0\). Then \(R\icx\rho\) for every uniformly integrable embedding \(\rho\) of \(\mu_1\) from \(\mu_0\) satisfying \(\E[\rho]<\infty\).
\end{lemma}

\begin{proof}[Proof of \Cref{lem:root-icx}]
If \(\mu_0=\mu_1\), take \(R=0\). Since every embedding time is nonnegative, \(R\icx\rho\), so the conclusion follows. Hence, suppose \(\mu_0\neq\mu_1\). \citet[Theorem 3.3]{CoxOblojTouzi2019}, specialized to one marginal, states that the Root embedding minimizes \(\E[\int_0^\rho f(s)\dd s]\) over all uniformly integrable embeddings whenever \(f\) is nonnegative and nondecreasing. Fix \(r\geq0\). Let \((f_{r,n})_{n\geq1}\) be nonnegative continuous nondecreasing functions satisfying \(f_{r,n}(s)\uparrow\1_{\{s>r\}}\) for every \(s\neq r\). \citet[Theorem 3.3]{CoxOblojTouzi2019} produces \(\E[\int_0^R f_{r,n}(s)\dd s]\leq\E[\int_0^\rho f_{r,n}(s)\dd s]\).
The monotone convergence theorem yields \(\E[(R-r)^+]\leq\E[(\rho-r)^+]\), and since this holds for every \(r\geq0\), \(R\icx\rho\).
\end{proof}

\begin{proof}[Proof of \Cref{thm:clock-bgng}]
Fix the disclosure clock \(C\) in the theorem, retain the corresponding filtration \(\mathbb J\) and full-speed posterior \(Q\), and set \(U\coloneqq C_\tau\). By \Cref{def:clock-domination}, \(U\) is a stopping time for \(Q\) and \(Q_U=\widehat P_\tau\). Clock domination implies \(U\leq\tau\), so \(U\) is integrable. Because \(Q\in[0,1]\), \((Q_{u\wedge U})_{u\geq0}\) is uniformly integrable. We proceed through three claims.

\begin{claim}\label{claim:posterior-diffusion-root-conditions}
The shifted full-speed posterior satisfies the hypotheses of \Cref{lem:root-icx}.
\end{claim}

Shift the posterior by setting \(\widetilde Q_u\coloneqq Q_u-p_0\). Its state interval is \(\mathcal I=(-p_0,1-p_0)\), and it satisfies
\[
\dd\widetilde Q_u=\eta(\widetilde Q_u)\dd W_u,
\qquad
\text{for }\eta(x)\coloneqq\kappa(x+p_0)(1-p_0-x).
\]
The coefficient \(\eta\) is locally Lipschitz, strictly positive in the interior of \(\mathcal I\), and bounded. The endpoints are inaccessible. Indeed, by It\^o's formula,
\[
\dd\left(\log\frac{Q_u}{1-Q_u}\right)=\kappa\dd W_u+\frac{\kappa^2}{2}(2Q_u-1)\dd u.
\]
The drift is bounded and the volatility is constant, so the log odds cannot explode at a finite time.

The processes \(P\) and \(Q\) solve the same stochastic differential equation from \(p_0\), so they have the same law. Under \(\Pp^\ell\),
\[
\log\frac{P_u}{1-P_u}=\log\frac{p_0}{1-p_0}+\kappa B_u^\ell-\frac{\kappa^2}{2}u.
\]
The strong law of large numbers for Brownian motion implies \(B_u^\ell/u\to0\) almost surely, so \(P_u\to0\) under \(\Pp^\ell\). Under \(\Pp^h\), \(B_u^\ell=B_u^h+\kappa u\), and hence \(P_u\to1\). Therefore, \(\widetilde Q\) approaches each finite inaccessible endpoint with positive probability as \(u\to\infty\).

\begin{claim}\label{claim:root-reduces-information-time-dispersion}
Root stopping produces \(\nu\) in the full-speed posterior \(Q\) at a time \(R_Q^\nu\preceq_{\mathrm{cx}}U\).
\end{claim}
Since \(U\) is an integrable stopping time that embeds \(\nu\) in \(Q\), \Cref{prop:root-posterior-timing} delivers a deterministic Root barrier \(\mathcal B^\nu\) such that, setting \(R_Q^\nu\coloneqq R_{\mathcal B^\nu}(Q)\), \(R_Q^\nu\) is integrable, \(\Law(Q_{R_Q^\nu})=\nu\), and \(R_Q^\nu\preceq_{\mathrm{cx}}U\).

\begin{claim}\label{claim:root-barrier-transfer-to-transparent-posterior}
The same deterministic barrier produces a transparent stopping time \(R^\nu\) with the required terminal distribution and timing law.
\end{claim}

Use the same barrier for the fully informed posterior and set \(R^\nu\coloneqq R_{\mathcal B^\nu}(P)\). Since \(P\) and \(Q\) solve the same stochastic differential equation from the same initial condition, \(\Law(R^\nu,P_{R^\nu})=\Law(R_Q^\nu,Q_{R_Q^\nu})\). Thus, \(\Law(P_{R^\nu})=\nu\) and \(R^\nu\preceq_{\mathrm{cx}}C_\tau\). Clock domination implies \(C_\tau\leq\tau\) almost surely, and so \(R^\nu\icx\tau\).

If \(\E[c(\tau)]=\infty\), boundedness of \(F\) makes the original payoff \(-\infty\), so the payoff inequality is immediate. Otherwise, \(\E[c(R^\nu)]\leq\E[c(C_\tau)]\leq\E[c(\tau)]\), while equality of the terminal posterior distributions implies \(\E[F(P_{R^\nu})]=\E[F(\widehat P_\tau)]\).

Full transparency is clock-dominated with disclosure clock \(C_t=t\), so \(V^{\mathrm{garb}}_{F,c}\geq V^0_{F,c}\). Conversely, the transparent replacement of every policy in the domain of \(V^{\mathrm{garb}}_{F,c}\) lies in the domain of \(V^0_{F,c}\) and yields higher expected payoff. Taking suprema yields \(V^{\mathrm{garb}}_{F,c}\leq V^0_{F,c}\).

If \(\tau\leq T\), then \(C_\tau\leq T\). Convex order delivers \(\E[(R^\nu-T)^+]\leq\E[(C_\tau-T)^+]=0\), and hence \(R^\nu\leq T\) almost surely. If \(c\) is strictly increasing and \(\Pp(C_\tau<\tau)>0\), then \(c(C_\tau)\leq c(\tau)\) almost surely, with strict inequality on a set of positive probability. Therefore, \(\E[c(R^\nu)]\leq\E[c(C_\tau)]<\E[c(\tau)]\). Finally, if \(c\) is strictly convex and \(\Law(R^\nu)\neq\Law(C_\tau)\), strict Jensen's inequality under a martingale coupling for \(R^\nu\preceq_{\mathrm{cx}}C_\tau\) yields \(\E[c(R^\nu)]<\E[c(C_\tau)]\).
\end{proof}

\subsection{Proof of \texorpdfstring{\Cref{ex:fast-binary-summary}}{ex}}\label[appendix]{app:exproof}
\begin{proof}[Proof of \Cref{ex:fast-binary-summary}]
Set \(p_0=1/2\), and choose \(\kappa>0\) so that \(\Phi(\kappa/2)=2/3\), where \(\Phi\) is the standard normal distribution function. Write \(L_t\coloneqq\log(P_t/(1-P_t))\). By \eqref{eq:samefilt}, conditional on the high and low states, respectively,
\[
L_1\sim\mathcal N\left(\frac{\kappa^2}{2},\kappa^2\right)
\qquad\text{and}\qquad
L_1\sim\mathcal N\left(-\frac{\kappa^2}{2},\kappa^2\right).
\]
Define \(D\coloneqq\1_{\{L_1\geq0\}}\). Our choice of \(\kappa\) implies
\[
\Pp(D=1)=\frac12,
\qquad
\Pp(\mu=\mu_h\mid D=1)=\frac23,
\qquad \text{and} \qquad
\Pp(\mu=\mu_h\mid D=0)=\frac13.
\]

Let \(H^\ast\coloneqq\inf\{t\geq0\colon P_t=2/3\}\) and \(A^\ast\coloneqq\{H^\ast\leq1\}\). Since \(A^\ast\in\calF_{H^\ast\wedge1}\) and \(P_{H^\ast\wedge1}=2/3\) on \(A^\ast\), optional sampling implies
\[
\Pp(\mu=\mu_h,A^\ast)
=
\E\left[P_{H^\ast\wedge1}\1_{A^\ast}\right]
=
\frac23\Pp(A^\ast).
\]
Because \(p_0=1/2\), this is equivalent to
\(\Pp^h(A^\ast)=2\Pp^\ell(A^\ast)\). Therefore, using
\(\{Z_1\geq1\}=\{D=1\}\),
\[
\Pp(A^\ast)
=
\frac32\E^\ell[(Z_1-1)\1_{A^\ast}]
<
\frac32\E^\ell[(Z_1-1)^+]
=
\frac32\left(\Pp^h(D=1)-\Pp^\ell(D=1)\right)
=
\frac12.
\]
The inequality is strict because
\(A^\ast\cap\{Z_1<1\}\) has positive \(\Pp^\ell\)-probability: \(L\) can reach \(\log2\) before date one and end below zero. Thus, \(\Pp(H^\ast\leq1)<\frac12\).

For \(p\in(1/2,2/3]\), let \(H_p\coloneqq\inf\{t\geq0\colon P_t=p\}\). As \(p\uparrow2/3\), continuity of \(P\) implies \(\{H_p\leq1\}\downarrow\{H^\ast\leq1\}\). Fix \(\bar P\in(1/2,2/3)\) sufficiently close to \(2/3\) that, writing \(H\coloneqq H_{\bar P}\), \(\Pp(H\leq1)<1/2\). The favorable realization of \(D\) now raises the posterior strictly above the persuasion threshold, while the unfavorable realization lowers it strictly below \(1-\bar P\).

Now let \(V\) be a Brownian motion independent of the state and the evidence. For \(\lambda>0\), define the public report
\[
Y_t^\lambda
\coloneqq
\lambda(2D-1)(t-1)^++V_t-V_{t\wedge1}.
\]
The report is silent through date one. Thereafter, its drift is \(\lambda\) on \(\{D=1\}\) and \(-\lambda\) on \(\{D=0\}\), so its noisy increments gradually reveal whether \(L_1\) was nonnegative. It is admissible under the representation in \eqref{eq:primitive-report}: its finite-variation term depends only on evidence available at date one, and its martingale term uses state-independent Gaussian noise. Let \(\G^\lambda\) be the usual augmentation of its natural filtration, let \(\widehat P^\lambda\) be the public posterior, and set
\(\tau_\lambda
\coloneqq
\inf\left\{t\geq1\colon
\widehat P_t^\lambda\notin(1-\bar P,\bar P)
\right\}\).

Write \(\pi_t^\lambda\coloneqq\Pp(D=1\mid\calG_t^\lambda)\). Because the report depends on the state only through \(D\), \(\widehat P_t^\lambda=1/3+\pi_t^\lambda/3\). The two stopping boundaries correspond to \(\pi_-\coloneqq2-3\bar P\) and \(\pi_+\coloneqq3\bar P-1=1-\pi_-\), which lie strictly between zero and one. Set \(b\coloneqq\log(\pi_+/\pi_-)\). For \(t\geq1\), the public log odds about \(D\) satisfy
\[
\log\frac{\pi_t^\lambda}{1-\pi_t^\lambda}
=
2\lambda\left(Y_t^\lambda-Y_1^\lambda\right).
\]
Conditional on \(D=1\), the public log odds \(s\) units after date one equal \(2\lambda^2s+2\lambda(V_{1+s}-V_1)\); and conditional on \(D=0\), they have the reflected law. Under the time change \(u=\lambda^2s\), Brownian scaling makes the first process equal in law to \(2u+2\widetilde B_u\), where \(\widetilde B\) is a standard Brownian motion. Let \(S\) be its first exit time from \((-b,b)\). Reflection symmetry delivers the same exit-time law conditional on \(D=0\), and so
\(\tau_\lambda\stackrel{d}{=}1+\frac{S}{\lambda^2}\). The random variable \(S\) is integrable because it is the first exit time of a Brownian motion with constant drift from a bounded interval. Since \(\widehat P^\lambda\) is a bounded martingale that starts at \(1/2\) and stops at \(1-\bar P\) or \(\bar P\), the optional sampling theorem implies \(1/2=q_\lambda\bar P+(1-q_\lambda)(1-\bar P)\), where \(q_\lambda\coloneqq\Pp(\widehat P_{\tau_\lambda}^\lambda=\bar P)\). Hence, \(q_\lambda=1/2\).

The public likelihood is continuous and strictly positive and so \Cref{prop:inverse-clock} supplies its canonical disclosure clock \(I^\lambda\). The filtering equation for the binary channel is
\[
\dd\pi_t^\lambda
=
2\lambda\pi_t^\lambda(1-\pi_t^\lambda)\dd W_t^\lambda,
\qquad \text{with } t>1,
\]
for a \(\G^\lambda\)-Brownian motion \(W^\lambda\). Its volatility is strictly positive before \(\tau_\lambda\). Hence, \(I_1^\lambda=0\), and \(I^\lambda\) is continuous and strictly increasing from date one until stopping.

Let \(U\) be the first exit time of a full-speed posterior from \((1-\bar P,\bar P)\). In the representation supplied by \Cref{prop:inverse-clock}, the range of \(I^\lambda\) before stopping is \([0,I_{\tau_\lambda}^\lambda)\), the public posterior remains inside \((1-\bar P,\bar P)\) before \(\tau_\lambda\), and it reaches the boundary at \(\tau_\lambda\). Accordingly, \(I_{\tau_\lambda}^\lambda\stackrel{d}{=}U\). The random variable \(U\) has unbounded support because a full-speed posterior has positive probability of remaining in the interval through every finite date.

Choose \(x>1\) with \(\Pp(U>x)>0\). Since \(\tau_\lambda\stackrel{d}{=}1+S/\lambda^2\),
\[
\Pp(\tau_\lambda>x)
=
\Pp\left(S>\lambda^2(x-1)\right)
\rightarrow0.
\]
For all sufficiently large \(\lambda\), the experiment cannot have any disclosure clock satisfying the clock bound. Otherwise, if \(C\) were such a clock and \(Q\) the corresponding full-speed posterior, the first exit time \(\widetilde U\) of \(Q\) from \((1-\bar P,\bar P)\) would satisfy \(\widetilde U\leq C_{\tau_\lambda}\leq\tau_\lambda\). Since \(\widetilde U\stackrel{d}{=}U\), this would imply \(\Pp(U>x)\leq\Pp(\tau_\lambda>x)\), a contradiction. The experiment, therefore, admits a disclosure clock but is not clock-dominated.

Finally, choose \(T>1\) such that \(\Pp(H\leq T)<1/2\), and then choose \(M>0\) large enough that \(q_T\coloneqq\Pp(H\leq T+1/M)<1/2\). Consider the delay cost \(c(t)\coloneqq M(t-T)^+\), which is finite, continuous, nondecreasing, and convex. Replacing a transparent stopping rule \(\rho\) by \(\rho\wedge H\) raises its persuasion probability and lowers its delay cost. We may, thus, suppose that \(\rho\leq H\). For any such rule,
\[
\1_{\{P_\rho\geq\bar P\}}-c(\rho)
=
\1_{\{\rho=H\}}-M(\rho-T)^+
\leq
\1_{\{H\leq T+1/M\}}.
\]
Consequently, every transparent stopping rule has payoff at most \(q_T\).

The policy \((\G^\lambda,\tau_\lambda)\) persuades with probability \(1/2\). Moreover, since \(T>1\),
\[
\E[c(\tau_\lambda)]
\leq
\frac{M}{\lambda^2}\E[S]
\rightarrow0.
\]
Choose \(\lambda\) large enough that the experiment is not clock-dominated and
\[
\frac12-\frac{M}{\lambda^2}\E[S]>q_T.
\]
The resulting garbling strictly outperforms every transparent stopping rule.
\end{proof}

\subsection{Proof of \texorpdfstring{\Cref{lem:threshold-normalization}}{the lemma}}\label[appendix]{app:proof-threshold-normalization}

\begin{proof}[Proof of \Cref{lem:threshold-normalization}]
Continuity of \(P\) and \(p_0<\bar P\) imply \(\1_{\{P_\rho\geq\bar P\}}\leq\1_{\{H\leq\rho\}}=\1_{\{P_{\rho\wedge H}\geq\bar P\}}\). Moreover, \(\rho\wedge H\leq\rho\), so by monotonicity of \(c\), \(c(\rho\wedge H)\leq c(\rho)\). Suppose \(\rho\) is transparent-optimal. The payoff gain from replacing \(\rho\) by \(\rho\wedge H\) is the expectation of \(\1_{\{H\leq\rho\}}-\1_{\{P_\rho\geq\bar P\}}+c(\rho)-c(\rho\wedge H)\), which is nonnegative almost surely. Optimality forces this expectation to equal zero. This random variable, thus, vanishes almost surely, so the persuasion indicators coincide. Thus, \(\rho\wedge H\) is optimal and preserves the persuasion probability.
\end{proof}

\subsection{Proof of \texorpdfstring{\Cref{lem:transparent-optimal-stopping-lattice}}{the lemma}}\label[appendix]{app:proof-transparent-optimal-stopping-lattice}

\begin{proof}[Proof of \Cref{lem:transparent-optimal-stopping-lattice}]
We proceed through four claims.

\begin{claim}\label{claim:finite-horizon-optimal-stopping-lattice}
For every finite horizon \(n\), an optimal stopping rule exists; the finite-horizon optimizer set is closed under pairwise minimum and maximum and has a least element \(\rho_n\).
\end{claim}

Define the adapted reward process \(G_t\coloneqq\1_{\{H\leq t\}}-c(t)\). For each integer \(n\geq1\), let \(V_n\coloneqq\sup_{\rho\leq n}\E[G_\rho]\), where the supremum is over \(\F^P\)-stopping times. The sequence \((V_n)_{n\geq1}\) is nondecreasing and bounded above by one. On \([0,n]\), the process \(G\) is bounded and c\`adl\`ag. If stopping times \(\theta_k\uparrow\theta\leq n\), continuity of \(c\) and the fact that \(t\mapsto\1_{\{H\leq t\}}\) has only an upward jump imply \(\limsup_{k\to\infty}G_{\theta_k}\leq G_\theta\) almost surely.
If \(\theta_k\downarrow\theta\), right-continuity implies \(G_{\theta_k}\to G_\theta\) almost surely. Boundedness turns both pathwise statements into the corresponding statements in expectation. Consequently, the finite-horizon Snell-envelope theorem yields an optimal stopping time. The set of horizon-\(n\) optimizers is closed under pairwise minimum and maximum. Indeed, if \(\theta\) and \(\eta\) are horizon-\(n\) optimal, then \(G_{\theta\wedge\eta}+G_{\theta\vee\eta}=G_\theta+G_\eta\) pathwise. Each expected payoff on the left is at most \(V_n\), while their sum equals \(2V_n\), so both stopping times are optimal. Because the optimizer set is closed under pairwise minima, there is a decreasing sequence of horizon-\(n\) optimizers whose limit is its essential infimum, namely the largest random time that is almost surely no greater than every horizon-\(n\) optimizer. Right-continuity of \(\F^P\) makes the limit a stopping time, and right-continuity and boundedness of \(G\) imply optimality by the bounded convergence theorem. Let \(\rho_n\) denote this least horizon-\(n\) optimizer.

\begin{claim}\label{claim:finite-horizon-optimizers-converge}
The least finite-horizon optimizers are increasing in the horizon, and their limit is an integrable global optimizer.
\end{claim}

The stopping time \(\rho_n\wedge H\) dominates \(\rho_n\) in the horizon-\(n\) problem and so is also optimal. Minimality of \(\rho_n\) implies \(\rho_n\leq\rho_n\wedge H\leq\rho_n\), so \(\rho_n\leq H\) almost surely.

The least finite-horizon optimizers are increasing in the horizon. Fix \(m>n\), and set \(a\coloneqq\rho_n\wedge\rho_m\) and \(b\coloneqq\rho_n\vee\rho_m\).
Then \(a\leq n\), \(b\leq m\), and the unordered pair \(\{a,b\}\) equals \(\{\rho_n,\rho_m\}\) pathwise. Hence, \(\E[G_a]+\E[G_b]=V_n+V_m\).
Feasibility implies \(\E[G_a]\leq V_n\) and \(\E[G_b]\leq V_m\), so both inequalities are equalities. Thus, \(a\) is horizon-\(n\) optimal. Since \(\rho_n\) is the least horizon-\(n\) optimizer, \(\rho_n\leq a\leq\rho_n\), and so \(\rho_n\leq\rho_m\) almost surely.

Set \(\rho^\ast\coloneqq\lim_{n\to\infty}\rho_n\). Immediate stopping yields zero, so \(V_n\geq0\). Since \(\rho_n\leq H\), \(0\leq V_n=\Pp(\rho_n=H)-\E[c(\rho_n)]\), and so \(\E[c(\rho_n)]\leq1\). The monotone convergence theorem implies \(\E[c(\rho^\ast)]\leq1\).
Because \(c\) is nonzero, choose \(t_0>0\) with \(c(t_0)>0\). Convexity and \(c(0)=0\) imply \(c(t)\geq\frac{c(t_0)}{t_0}t\) for every \(t\geq t_0\), and hence \(t\leq t_0+\frac{t_0}{c(t_0)}c(t)\) for every \(t\geq0\). Thus, \(\E[\rho^\ast]<\infty\), so \(\rho^\ast\in\mathcal T_H\).

The events \(\{\rho_n=H\}\) are increasing, and their union is contained in \(\{\rho^\ast=H\}\). Hence, \(J_c(\rho^\ast)\geq\lim_{n\to\infty}V_n\).
To identify the limit, fix \(\rho\in\mathcal T_H\). Then \(\rho\wedge n\leq n\), \(\1_{\{\rho\wedge n=H\}} \rightarrow \1_{\{\rho=H\}}\) almost surely, and \(c(\rho\wedge n)\uparrow c(\rho)\). Therefore, \(J_c(\rho\wedge n) \rightarrow  J_c(\rho)\), including when the limit equals \(-\infty\). Since \(V_n\geq J_c(\rho\wedge n)\), taking limits and then the supremum over \(\rho\in\mathcal T_H\) yields \(\lim_nV_n\geq\mathcal V(c)\). The reverse inequality follows from \(\rho_n\in\mathcal T_H\). Thus, \(\lim_nV_n=\mathcal V(c)\), and feasibility of \(\rho^\ast\) forces \(J_c(\rho^\ast)=\mathcal V(c)\). Therefore, \(\mathcal S(c)\) is nonempty.

\begin{claim}\label{claim:global-optimizer-lattice}
The global optimizer set \(\mathcal S(c)\) is closed under pairwise minimum and maximum.
\end{claim}

For \(\rho,\eta\in\mathcal T_H\), the pair \(\{\rho\wedge\eta,\rho\vee\eta\}\) equals \(\{\rho,\eta\}\) pathwise. Since \(\rho,\eta\leq H\), this implies \(\1_{\{\rho\wedge\eta=H\}}+\1_{\{\rho\vee\eta=H\}}=\1_{\{\rho=H\}}+\1_{\{\eta=H\}}\).
Moreover, \(c(\rho\wedge\eta)+c(\rho\vee\eta)=c(\rho)+c(\eta)\), and hence \(J_c(\rho\wedge\eta)+J_c(\rho\vee\eta)=J_c(\rho)+J_c(\eta)\).
If \(\rho,\eta\in\mathcal S(c)\), each term on the left is at most \(\mathcal V(c)\), while their sum equals \(2\mathcal V(c)\). Thus, both \(\rho\wedge\eta\) and \(\rho\vee\eta\) belong to \(\mathcal S(c)\).

\begin{claim}\label{claim:extremal-global-optimizers}
The set \(\mathcal S(c)\) contains its essential infimum and essential supremum, which are its least and greatest elements.
\end{claim}

Because \(\mathcal S(c)\) is downward directed, there is a decreasing sequence \((\sigma_n)_{n\geq1}\subseteq\mathcal S(c)\) such that \(\sigma_n\downarrow\rho_c^-\coloneqq\operatorname*{ess\,inf}_{\rho\in\mathcal S(c)}\rho\).
The filtration \(\F^P\) is right-continuous, so \(\rho_c^-\) is a stopping time. Moreover, \(\rho_c^-\leq\sigma_1\), so \(\E[\rho_c^-]<\infty\). Since every optimizer has finite expected cost and \(c(\sigma_n)\leq c(\sigma_1)\), the dominated convergence theorem yields \(\E[c(\sigma_n)] \rightarrow \E[c(\rho_c^-)]\). Because \(\sigma_n\leq H\), \(\1_{\{\sigma_n=H\}} \rightarrow \1_{\{\rho_c^-=H\}}\) almost surely.
Therefore, \(J_c(\rho_c^-)=\mathcal V(c)\), so \(\rho_c^-\in\mathcal S(c)\).

Likewise, upward directedness produces an increasing sequence \((\eta_n)_{n\geq1}\subseteq\mathcal S(c)\) such that \(\eta_n\uparrow\rho_c^+\coloneqq\operatorname*{ess\,sup}_{\rho\in\mathcal S(c)}\rho\). Optimality implies \(\E[c(\eta_n)]=\Pp(\eta_n=H)-\mathcal V(c)\leq1\).
The monotone convergence theorem and the eventual linear lower bound on \(c\) imply \(\E[\rho_c^+]<\infty\). Moreover, \(\1_{\{\eta_n=H\}}\uparrow\1_{\bigcup_n\{\eta_n=H\}}\leq\1_{\{\rho_c^+=H\}}\).
Hence, \(J_c(\rho_c^+)\geq\mathcal V(c)\). Feasibility delivers the reverse inequality, so \(\rho_c^+\in\mathcal S(c)\). By construction, \(\rho_c^-\) and \(\rho_c^+\) are respectively the least and greatest elements of \(\mathcal S(c)\).
\end{proof}

\subsection{Proof of \texorpdfstring{\Cref{prop:higher-delay-cost-lower-persuasion}}{the proposition}}\label[appendix]{app:proof-higher-delay-cost-lower-persuasion}

\begin{proof}[Proof of \Cref{prop:higher-delay-cost-lower-persuasion}]
Let \(\Delta\coloneqq c_2-c_1\). Since \(\Delta\) is nondecreasing and \(\Delta(0)=0\), \(c_1\leq c_2\). Fix \(\rho_1\in\mathcal S(c_1)\) and \(\rho_2\in\mathcal S(c_2)\), and set \(m\coloneqq\rho_1\wedge\rho_2\) and \(j\coloneqq\rho_1\vee\rho_2\).
Every optimizer has finite expected cost. Since \(c_1\leq c_2\), \(\E[c_1(\rho_2)]<\infty\). Moreover, \(c_2(m)\leq c_2(\rho_2)\) and \(c_1(j)\leq c_1(\rho_1)+c_1(\rho_2)\), so all payoff differences below are finite.

Because \(m\leq\rho_2\), monotonicity of \(\Delta\) and optimality of \(\rho_2\) under \(c_2\) deliver
\[
J_{c_1}(\rho_2)-J_{c_1}(m)=J_{c_2}(\rho_2)-J_{c_2}(m)+\E[\Delta(\rho_2)-\Delta(m)]\geq0.
\]
The modular identiy established in \Cref{lem:transparent-optimal-stopping-lattice}, applied under \(c_1\), yields \(J_{c_1}(j)-J_{c_1}(\rho_1)=J_{c_1}(\rho_2)-J_{c_1}(m)\geq0\).
Since \(\rho_1\) is \(c_1\)-optimal, the left-hand side cannot be positive. Hence, equality holds and \(j\in\mathcal S(c_1)\). Moreover, \(J_{c_1}(\rho_2)-J_{c_1}(m)\) is the sum of two nonnegative terms and equals zero, so \(J_{c_2}(\rho_2)=J_{c_2}(m)\).
Thus, \(m\in\mathcal S(c_2)\), proving the strong set order.

Apply the strong set order to \(\rho_{c_1}^-\) and \(\rho_{c_2}^-\). Their minimum belongs to \(\mathcal S(c_2)\), and minimality of \(\rho_{c_2}^-\) implies \(\rho_{c_2}^-\leq\rho_{c_1}^-\wedge\rho_{c_2}^-\leq\rho_{c_2}^-\).
Therefore, \(\rho_{c_2}^-\leq\rho_{c_1}^-\). The strong set order also implies \(\rho_{c_1}^+\vee\rho_{c_2}^+\in\mathcal S(c_1)\). Since \(\rho_{c_1}^+\) is the greatest \(c_1\)-optimizer, \(\rho_{c_1}^+\leq\rho_{c_1}^+\vee\rho_{c_2}^+\leq\rho_{c_1}^+\), and so \(\rho_{c_2}^+\leq\rho_{c_1}^+\). The persuasion-probability inequalities follow because \(\{\rho=H\}\) is increasing in \(\rho\) on \(\mathcal T_H\).
\end{proof}

\subsection{Proof of \texorpdfstring{\Cref{prop:higher-snr-higher-persuasion}}{the proposition}}\label[appendix]{app:proof-higher-snr-higher-persuasion}

\begin{proof}[Proof of \Cref{prop:higher-snr-higher-persuasion}]
Set \(a\coloneqq\frac{1}{\kappa_1^2}\) and \(b\coloneqq\frac{1}{\kappa_2^2}\), so \(a\geq b>0\). The cost difference is \(d_{\kappa_1}(u)-d_{\kappa_2}(u)=c(au)-c(bu)\).
A finite convex function is locally absolutely continuous, and its right derivative \(c_+'\) is nonnegative and nondecreasing. For \(v>u\),
\[
[c(av)-c(bv)]-[c(au)-c(bu)]=\int_u^v[a c_+'(as)-b c_+'(bs)]\dd s\geq0,
\]
because \(as\geq bs\), \(c_+'(as)\geq c_+'(bs)\geq0\), and \(a\geq b\). Hence, \(d_{\kappa_1}-d_{\kappa_2}\) is nondecreasing.

Apply \Cref{prop:higher-delay-cost-lower-persuasion} to the common process \(P^1\), with \(d_{\kappa_2}\) as the lower cost and \(d_{\kappa_1}\) as the higher cost. This yields the strong set order comparison and the inequalities for the extremal stopping rules. The persuasion-probability inequalities follow from monotonicity of \(\{U=H_1\}\) in \(U\) on \(\mathcal T_{H_1}^1\).
\end{proof}

\subsection{Proof of \texorpdfstring{\Cref{lem:receiver-operational-pullback}}{the lemma}}\label[appendix]{app:proof-receiver-operational-pullback}

\begin{proof}[Proof of \Cref{lem:receiver-operational-pullback}]
We proceed through three claims.

\begin{claim}\label{claim:receiver-clock-pullback-identity}
For every \(u\geq0\), \(C_{\theta_u}=u\wedge U\), \(\widehat P_{\theta_u}^{\mathsf S}=Q_{u\wedge U}\), and \(U\) is an \(\mathbb H\)-stopping time.
\end{claim}

Continuity and monotonicity of \(C\) imply that \(T_u^+\) is a \(\mathbb G^{\mathsf S}\)-stopping time. On \(\{U\leq u\}\), the clock does not move strictly above \(u\) before \(\rho\), so \(\theta_u=\rho\) and \(C_{\theta_u}=U\). On \(\{U>u\}\), continuity implies \(T_u^+<\rho\) and \(C_{T_u^+}=u\). Hence, \(C_{\theta_u}=u\wedge U\). The equation in \Cref{def:clock-domination} then yields \(\widehat P_{\theta_u}^{\mathsf S}=Q_{u\wedge U}\).

Moreover, \(\{U\leq u\}=\{\rho\leq T_u^+\}\). For two stopping times, the event that one precedes the other belongs to the sigma-field at their minimum. Therefore, \(\{U\leq u\}\in\calG_{\theta_u}^{\mathsf S}=\calH_u^0\subseteq\calH_u\), proving that \(U\) is an \(\mathbb H\)-stopping time.

\begin{claim}\label{claim:receiver-calendar-pullback-stopping-time}
If \(V\leq U\) is an \(\mathbb H\)-stopping time, then \(\theta_V\) is a \(\mathbb G^{\mathsf S}\)-stopping time.
\end{claim}

Let \(V\leq U\) be an \(\mathbb H\)-stopping time. For \(n\geq1\), define the upper dyadic approximation \(V_n\coloneqq2^{-n}(\lfloor2^nV\rfloor+1)\). Then \(V_n\downarrow V\). Right-continuity of \(\mathbb H\) makes every \(V_n\) an \(\mathbb H\)-stopping time. For each \(n\), the random time \(V_n\) has countable range. Since \(\mathbb H\) is the completion of \(\mathbb H^0\), after modifying on one null set we may take each event \(\{V_n=u\}\) in \(\calH_u^0=\calG_{\theta_u}^{\mathsf S}\). Pasting the stopping times \(\theta_u\) on these events shows that \(\theta_{V_n}\) is a \(\mathbb G^{\mathsf S}\)-stopping time. The generalized inverse \(u\mapsto T_u^+\) is right-continuous because \(C\) is continuous and nondecreasing. Thus, \(\theta_{V_n}\downarrow\theta_V\). Right-continuity of \(\mathbb G^{\mathsf S}\) implies that \(\theta_V\) is a stopping time.

\begin{claim}\label{claim:receiver-pullback-payoff-comparison}
The identities \(\widehat P_{\theta_V}^{\mathsf S}=Q_V\) and \(\theta_V-V\leq\rho-U\) hold almost surely.
\end{claim}

On the full-probability event on which \(C\) is continuous and unbanked, \(C_{\theta_u}=u\wedge U\) holds for every \(u\geq0\). Evaluating it at \(V\) yields \(C_{\theta_V}=V\) and \(\widehat P_{\theta_V}^{\mathsf S}=Q_V\).

On \(\{V<U\}\), \(\theta_V<\rho\). Unbankedness applied between \(\theta_V\) and \(\rho\) delivers \(U-V=C_\rho-C_{\theta_V}\leq\rho-\theta_V\). Rearranging yields \(\theta_V-V\leq\rho-U\). On \(\{V=U\}\), \(\theta_V=\rho\) by definition, so equality holds. Since \(V\leq U\), the same inequality also implies \(\theta_V\leq\rho\); in particular, \(\theta_V\) is integrable.
\end{proof}

\subsection{Proof of \texorpdfstring{\Cref{lem:receiver-stopped-experiment-transplant}}{the lemma}}\label[appendix]{app:proof-receiver-stopped-experiment-transplant}

\begin{proof}[Proof of \Cref{lem:receiver-stopped-experiment-transplant}]
Retain \(\mathbb J\) and \(Q\) fixed in \secref{sec:uncommitted-receiver}, and let \(\mathbb F^Q\) be the usual augmentation of the natural filtration of \(Q\).
We proceed through four claims.

\begin{claim}\label{claim:posterior-filtration-immersed}
For every \(t\geq0\) and \(B\in\mathcal F_\infty^Q\), \(\E[\1_B\mid\mathcal J_t]=\E[\1_B\mid\mathcal F_t^Q]\).
\end{claim}

The process \(Q\) is driven by a \(\mathbb J\)-Brownian motion \(W\). Since \(Q_u\in(0,1)\) at every finite date, \(W_t=\int_0^t\frac{1}{\kappa Q_s(1-Q_s)}\dd Q_s\), so the path of \(Q\) determines the path of \(W\). Thus, \(\mathbb F^Q\) is also the usual natural filtration of \(W\).

Fix \(B\in\mathcal F_\infty^Q\), and set \(M_t\coloneqq\E[\1_B\mid\mathcal F_t^Q]\). By the martingale representation theorem for Brownian motion, there is an \(\mathbb F^Q\)-predictable process \(h\) such that \(M_t=M_0+\int_0^th_s\dd W_s\). Since \(\mathbb F^Q\subseteq\mathbb J\), the process \(h\) is also \(\mathbb J\)-predictable. Because \(W\) is a \(\mathbb J\)-Brownian motion, \(M\) is a \(\mathbb J\)-martingale. Moreover, \(M_t\to\1_B\) in \(L^1\). Hence, \(M_t=\E[\1_B\mid\mathcal J_t]\), proving the claim.

\begin{claim}\label{claim:receiver-stopping-kernel}
There is a probability kernel \(K\) such that \(K(Q,\cdot)\) is a conditional distribution of \(U\) given \(\mathcal F_\infty^Q\) and \(K(Q,[0,t])\) is \(\mathcal F_t^Q\)-measurable for every \(t\geq0\).
\end{claim}

Continuous path space, equipped with uniform convergence on compact intervals, is a complete separable metric space, so a regular conditional distribution of \(U\) given \(\mathcal F_\infty^Q\) exists. This is a measurable map that assigns a probability distribution over stopping dates to each posterior path. Write it as \(K(Q,\dd s)\). For every \(t\geq0\), \(K(Q,[0,t])=\E[\1_{\{U\leq t\}}\mid\mathcal F_\infty^Q]\).

The event \(\{U\leq t\}\) belongs to \(\mathcal J_t\). Fix \(B\in\mathcal F_\infty^Q\). The tower property yields \(\E[\1_{\{U\leq t\}}\1_B]=\E[\1_{\{U\leq t\}}\E[\1_B\mid\mathcal J_t]]\). By \Cref{claim:posterior-filtration-immersed}, this equals \(\E[\1_{\{U\leq t\}}\E[\1_B\mid\mathcal F_t^Q]]\), which in turn equals \(\E[\E[\1_{\{U\leq t\}}\mid\mathcal F_t^Q]\1_B]\). Hence, \(\E[\1_{\{U\leq t\}}\mid\mathcal F_\infty^Q]=\E[\1_{\{U\leq t\}}\mid\mathcal F_t^Q]\), so \(K(Q,[0,t])\) is \(\mathcal F_t^Q\)-measurable. Choose the kernel so that this measurability holds simultaneously at every rational date, and use right-continuity of its distribution function to extend it to every date.

\begin{claim}\label{claim:transparent-randomized-stopping-time}
Applying the same kernel to \(P\) produces \(\sigma_D\) such that \(((P_u)_{u\geq0},\sigma_D)\stackrel{d}{=}((Q_u)_{u\geq0},U)\).
\end{claim}

By uniqueness in law for \eqref{eq:full-speed-posterior}, \(P\) and \(Q\) have the same distribution on continuous path space. Let \(\xi_D\) be uniform on \([0,1]\) and independent of \(\sigma(\mu,\calF_\infty)\), and define \(\sigma_D\coloneqq\inf\{t\geq0\colon K(P,[0,t])\geq\xi_D\}\). Since \(K(P,[0,t])\) is measurable from the path of \(P\) through \(t\), \(\sigma_D\) is a stopping time in the filtration generated by \(P\) and \(\xi_D\). Conditional on the path of \(P\), its law is \(K(P,\cdot)\). Therefore, \(((P_u)_{u\geq0},\sigma_D)\stackrel{d}{=}((Q_u)_{u\geq0},U)\).  Define
\[
Z^D\coloneqq
\left((P_{u\wedge\sigma_D})_{u\geq0},
(\1_{\{\sigma_D\leq u\}})_{u\geq0}\right)
 \qquad \text{and} \qquad
Z^Q\coloneqq
\left((Q_{u\wedge U})_{u\geq0},
(\1_{\{U\leq u\}})_{u\geq0}\right),
\]
so that we have \(Z^D\stackrel{d}{=}Z^Q\).

\begin{claim}\label{claim:receiver-stopping-rule-transfer}
Every stopping time \(\eta\leq\sigma_D\) for the usual natural filtration of \(Z^D\) has a stopping time \(V\leq U\) for the usual natural filtration of \(Z^Q\) such that \((P_\eta,\eta)\stackrel{d}{=}(Q_V,V)\).
\end{claim}

Regard \(Z^D\) and \(Z^Q\) as random elements of the same path space. Let \(\mathbb E^0=(\mathcal E_t^0)_{t\geq0}\) be the filtration generated directly by the coordinate paths, before completion or right-continuous augmentation, and let \(\mathbb E=(\mathcal E_t)_{t\geq0}\) be its right-continuous modification.

For each \(q\in\mathbb Q_+\), there is \(B_q\in\mathcal E_q^0\) such that \(\{\eta<q\}=\{Z^D\in B_q\}\) almost surely. Indeed, \(\{\eta<q\}=\bigcup_{r\in\mathbb Q_+,\,r<q}\{\eta\leq r\}\). For each \(r<q\), choose \(s\in(r,q)\). Because \(\eta\) is a stopping time for the usual augmentation of the natural filtration of \(Z^D\), the event \(\{\eta\leq r\}\) agrees almost surely with the inverse image of an event in \(\mathcal E_s^0\subseteq\mathcal E_q^0\). Taking the countable union produces \(B_q\).

Define \(A_q\coloneqq\bigcup_{r\in\mathbb Q_+,\,r<q}B_r\). Then \((A_q)_{q\in\mathbb Q_+}\) is increasing, \(A_q\in\mathcal E_q^0\), and \(\{Z^D\in A_q\}=\{\eta<q\}\) almost surely for every \(q\in\mathbb Q_+\). Since \(\mathbb Q_+\) is countable, these identities hold simultaneously outside one null set. Define \(\varphi(z)\coloneqq\inf\{q\in\mathbb Q_+\colon z\in A_q\}\), where \(\inf\varnothing=\infty\). For every \(t\geq0\), \(\{\varphi<t\}=\bigcup_{q\in\mathbb Q_+,\,q<t}A_q\in\mathcal E_t^0\). Right-continuity of \(\mathbb E\), thus, makes \(\varphi\) an \(\mathbb E\)-stopping time. For every \(q\in\mathbb Q_+\), density of \(\mathbb Q_+\) and the definition of \(A_q\) produce \(\{\varphi<q\}=A_q\). Hence, \(\varphi(Z^D)=\eta\) almost surely.

Writing a coordinate path as \(z=(x,d)\), define \(\zeta(z)\coloneqq\inf\{u\geq0\colon d_u=1\}\). \(\{\zeta\leq t\}=\{d_t=1\}\) makes \(\zeta\) an \(\mathbb E^0\)-stopping time. Replace \(\varphi\) by \(\varphi\wedge\zeta\). Since \(\eta\leq\sigma_D=\zeta(Z^D)\) almost surely, this replacement preserves \(\varphi(Z^D)=\eta\) almost surely and implies \(\varphi\leq\zeta\) on the path space. Set \(V\coloneqq\varphi(Z^Q)\). For every \(t\geq0\), the event \(\{V\leq t\}\) belongs to the usual natural filtration of \(Z^Q\) at \(t\), so \(V\) is a stopping time for that filtration. Moreover, \(\varphi\leq\zeta\) implies \(V\leq U\). Finally, \(Z^D\stackrel{d}{=}Z^Q\) and \(\varphi(Z^D)=\eta\) yield \((Z^D,\eta)\stackrel{d}{=}(Z^Q,V)\). Since the first-coordinate paths are continuous, evaluating them at the corresponding stopping times delivers \((P_\eta,\eta)\stackrel{d}{=}(Q_V,V)\).
\end{proof}

\subsection{Proof of \texorpdfstring{\Cref{thm:uncommitted-receiver-acceleration}}{the theorem}}\label[appendix]{app:proof-uncommitted-receiver-acceleration}

\begin{proof}[Proof of \Cref{thm:uncommitted-receiver-acceleration}]
We proceed through three claims.

\begin{claim}\label{claim:transparent-acceleration-admissible}
The transparent acceleration is clock-dominated and has an unbanked disclosure clock.
\end{claim}

Let \(\widetilde{\mathbb F}^P=(\widetilde{\calF}_u^P)_{u\geq0}\) be the usual augmentation of the filtration generated by \(P\) and the independent date-zero randomization used to construct \(\sigma_D\), and set \(C_t^D\coloneqq t\wedge\sigma_D\). Since the added randomization is independent of the state, \(P_u=\E[\1_{\{\mu=\mu_h\}}\mid\widetilde{\calF}_u^P]\). Optional sampling produces \(\E[\1_{\{\mu=\mu_h\}}\mid\widetilde{\calF}_{t\wedge\sigma_D}^P]=P_{t\wedge\sigma_D}\).

The raw public sigma-field \(\calG_t^{D,0}\) is contained in \(\widetilde{\calF}_{t\wedge\sigma_D}^P\), and the inclusion survives completion and right-continuous augmentation. Moreover, \(P_{t\wedge\sigma_D}\) is \(\calG_t^D\)-measurable by \Cref{def:transparent-acceleration}. As a result, the tower property implies
\[
\widehat P_t^D
=
\E\left[
\E\left[\1_{\{\mu=\mu_h\}}
\mid\widetilde{\calF}_{t\wedge\sigma_D}^P\right]
\middle|\calG_t^D
\right]
=
P_{t\wedge\sigma_D}
=
P_{C_t^D}.
\]
Thus, \(\calG_t^D\subseteq\widetilde{\calF}_{C_t^D}^P\). Each \(C_t^D\) is a \(\widetilde{\mathbb F}^P\)-stopping time. Moreover, if \(\tau\) is a \(\mathbb G^D\)-stopping time, then \(C_\tau^D=\tau\wedge\sigma_D\) is a \(\widetilde{\mathbb F}^P\)-stopping time. Hence, \(C^D\) is a disclosure clock. It satisfies, almost surely, \(C_t^D-C_s^D\leq t-s\) for every \(0\leq s\leq t\), so it is unbanked and, in particular, clock-dominated.

\begin{claim}\label{claim:transparent-acceleration-best-response}
The receiver optimally waits until \(\sigma_D\) under the transparent acceleration.
\end{claim}

By \Cref{lem:receiver-stopped-experiment-transplant}, \(\E[\sigma_D]=\E[U]\leq\E[\rho]<\infty\). Fix an integrable \(\mathbb G^D\)-stopping time \(\eta\). Since the public posterior is constant after \(\sigma_D\) and \(\kappa_R>0\), replacing \(\eta\) by \(\eta\wedge\sigma_D\) raises the receiver's payoff. Consequently, we  may suppose that \(\eta\leq\sigma_D\).

By \Cref{lem:receiver-stopped-experiment-transplant}, there is a stopping time \(V\leq U\) for the natural filtration of \(Z^Q\) such that \((P_\eta,\eta)\stackrel{d}{=}(Q_V,V)\). By \Cref{lem:receiver-operational-pullback}, both coordinate processes of \(Z^Q\) are \(\mathbb H\)-adapted, so \(V\) is an \(\mathbb H\)-stopping time. The same lemma makes \(\theta_V\) a \(\mathbb G^{\mathsf S}\)-stopping time, with \(\widehat P_{\theta_V}^{\mathsf S}=Q_V\) and \(\theta_V-V\leq\rho-U\). Optimality of \(\rho\) implies \(\E[r(Q_V)-\kappa_R\theta_V]\leq\E[r(Q_U)-\kappa_R\rho]\). Therefore,
\[
\begin{aligned}
\E[r(P_\eta)-\kappa_R\eta]
&=\E[r(Q_V)-\kappa_RV]\\
&=\E[r(Q_V)-\kappa_R\theta_V]+\kappa_R\E[\theta_V-V]\\
&\leq\E[r(Q_U)-\kappa_R\rho]+\kappa_R\E[\rho-U]\\
&=\E[r(Q_U)-\kappa_RU]
=\E[r(P_{\sigma_D})-\kappa_R\sigma_D].
\end{aligned}
\]
Thus, \(\sigma_D\) is a receiver best response to \(D\).

\begin{claim}\label{claim:transparent-acceleration-payoffs}
The transparent acceleration raises the sender's payoff and gives the receiver the stated gain.
\end{claim}

By \Cref{lem:receiver-stopped-experiment-transplant}, \(\Law(P_{\sigma_D})=\Law(Q_U)=\Law(\widehat P_\rho^{\mathsf S})\) and \(\Law(\sigma_D)=\Law(U)\). Since \(U=C_\rho\leq\rho\) and \(c\) is increasing, \(\E[c(\sigma_D)]=\E[c(U)]\leq\E[c(\rho)]\). Equality of the terminal posterior distributions proves the sender's payoff comparison. Finally,
\[
\E[r(P_{\sigma_D})-\kappa_R\sigma_D]
=
\E[r(\widehat P_\rho^{\mathsf S})-\kappa_R\rho]
+\kappa_R\E[\rho-C_\rho]. \qedhere
\]
\end{proof}
\subsection{Proof of \texorpdfstring{\Cref{cor:uncommitted-receiver-bgng}}{the corollary}}\label[appendix]{app:proof-uncommitted-receiver-bgng}

\begin{proof}[Proof of \Cref{cor:uncommitted-receiver-bgng}]
The conclusion follows from \Cref{thm:uncommitted-receiver-acceleration}.
\end{proof}

\end{document}